\documentstyle[emulateapj,psfig]{article}

\begin{document}

\newcommand{\be}{\begin{equation}}
\newcommand{\ee}{\end{equation}}     

\def\xr#1{\parindent=0.0cm\hangindent=1cm\hangafter=1\indent#1\par}
\def\ea{et al.}
\def\exo{{\sl EXOSAT}}
\def\ein{{\sl EINSTEIN }}
\def\gin{{\sl GINGA}}
\def\rosat{{\sl ROSAT}}
\def\asca{{\sl ASCA}}
\def\chandra{{\sl Chandra}}
\def\xmm{{\sl XMM-Newton}}
\def\as{$^{\prime\prime}$}
\def\am{$^{\prime}$}
\def\cps{{ ct~s$^{\rm -1}$ }}
\def\ergpsec{{ergs~s$^{\rm -1}$}}
\def\ergpcps{{ergs~cm$^{\rm -2}$~s$^{\rm -1}$}}
\def\g21{G21.5$-$0.9}

\def\shock{{\it PSHOCK }}
\def\nei{{\it NEI }}

\title{\bf X-Ray Observations of the supernova remnant G21.5$-$0.9}

\author{S. Safi-Harb \altaffilmark{1}}
\affil{University of Manitoba, Winnipeg, MB R3T 2N2, Canada;
 samar@physics.umanitoba.ca}

\author{I.~M. Harrus\altaffilmark{2}, R. Petre}
\affil{NASA's Goddard Space Flight Center, Greenbelt, MD 20771, USA}

\author{G.~G.~Pavlov, A.~B. Koptsevich \altaffilmark{3}, \& D. Sanwal}
\affil{Pennsylvania State University, University Park, PA, 16802, USA}

\altaffiltext{1} {NSERC UFA fellow.}
\altaffiltext{2} {Universities Space Research Association.}  
\altaffiltext{3} {On leave from Ioffe Physical
 Technical Institue, St. Petersburg, Russia.}

{\centerline{\it{submitted 2/13/01; accepted 06/22/01 for ApJ--Oct 20, 2001 issue}
}}

\begin{abstract}
We present the analysis of 
archival X-ray observations of the supernova remnant (SNR) \g21.
Based on its morphology and spectral properties,
\g21 has been classified as a Crab-like SNR.
For that reason, it was chosen as a \chandra \ calibration
target.
In their early analysis of part of these calibration data,
 Slane \ea \ (2000) discovered a low-surface-brightness, extended
emission.
They interpreted this component as the 
blast wave formed in the supernova (SN) explosion.
\xmm \ observations (Warwick \ea \ 2000)
 revealed the non-thermal nature of this
emission suggesting that it is instead an extension
of the synchrotron nebula. 
In this paper, we revisit the \chandra \ analysis
using new calibration data, improving
the statistics by a factor of 2.
We also include \rosat \ and \asca \ observations. 
Our analysis confirms the non-thermal nature of
the extended emission.
ACIS-S images indicate that this component is not limb-brightened,
 and it shows knotty structures and a bright filament
 2\am \ north of the center.
We find no evidence of line emission from any part of the remnant.
We can  reject a collisional
equilibrium ionization (CEI) thermal model at 
solar abundances, and non-equilibrium ionization (NEI) models
(such as a plane-parallel shock model with different ionization
ages and constant-temperature, or an NEI model with a single ionization age
and a constant temperature).
The entire remnant is best fitted
with a power law model with
a photon index steepening away from the center.
The total unabsorbed flux $F_X$ (0.5 -- 10 keV) is 1.1 $\times$
10$^{\rm -10}$~erg~cm$^{\rm -2}$~s$^{\rm -1}$ with an 85\%
contribution from the 40\as \ radius core.
Timing analysis of the High-Resolution Camera (HRC) data
failed to detect any pulsations.
We put a 16\% upper limit on the pulsed fraction.
We derive the physical parameters of the putative pulsar
and compare them with those of other plerions (such
as the Crab and 3C~58).


\end{abstract}

\section{Introduction}

Crab-like supernova remnants have a center-brightened
nebula, referred to as a plerion, often seen in the radio 
and X-ray wavelengths, and believed to be powered by an embedded pulsar.
Typified by the Crab nebula, they have non-thermal spectra
and lack the evidence of a SNR shell at all wavelengths.
 In the radio, they are highly
polarized and have a flat power law spectral index
($\alpha$ $\sim$ 0$-$0.3; $S_{\nu}$ $\sim$ $\nu^{-\alpha}$).
In X-rays, they have
a hard photon index ($\Gamma$ $\sim$ 2; where $\Gamma$ = $\alpha$+1).    
Out of the $\sim$ 220 Galactic supernova remnants (SNRs),
only $\sim$ 10 candidates are
classified as Crab-like supernova remnants (Green 2000). 

\g21 \ is a plerionic SNR.
In the radio, it has been imaged with the Very Large Array (VLA)
at 5~GHz (Becker and Szymkowiak, 1981), 
with the NRO Millimeter Array  at 22.3~GHz,
and with the Effelsberg 100-m telescope at 32 GHz
(Morsi \& Reich 1987). 
These observations show a 
centrally peaked highly polarized emission, of radius $\sim$ 40\as,
having a flat radio spectral index ($\alpha$~$\simeq$~0) 
and a flux density of 6~Jy at 1~GHz.
The 22.3~GHz map obtained with the NRO (F\"urst et al. 1988)
very closely resembles the 5~GHz map of Becker
and Szymkowiak (1981, Fig.~1), and exhibits an
ellipse of 1\am.5 major axis, an axial ratio of 0.8 and a position
angle $\sim$35$^\circ$.
21~cm absorption data obtained with the VLA
place the SNR at a distance between 4.8 and 9.3 kpc (Becker, \& Szymkowiak 1981;
Davelaar, Smith, \& Becker 1986). 
Assuming a distance of 5~kpc, the linear size
of the radio plerion is ~2.2~pc x 1.3~pc.
F\"urst et al. (1988) found, in addition to the
diffuse centrally peaked component, axisymmetric filaments
which they interpret as signatures of jets from a central pulsar.
Infrared observations of \g21 \   with ESA's
Infrared Space Observatory (ISOCAM; Gallant \& Tuffs 1998)
show that the plerion has a morphology 
similar to the 22~GHz map of F\"urst et al. (1988).

X-ray observations of \g21 point unambiguously to a centrally peaked 
non-thermal X-ray plerion.
\g21\ was observed with the non-imaging medium-energy (ME) instrument on-board \exo.
Analysis of these data indicates a power law spectrum with a photon index of
$\Gamma$~=~1.72~$\pm$~0.12, somewhat flatter
than that of the Crab (Davelaar \ea \ 1986).
A subsequent short observation (2816~s)
 by \gin\ in the 2--10~keV range could not distinguish
between a power law model with a photon index of $\Gamma$~=~1.88~$\pm$~0.05
and a thermal bremsstrahlung model with a very high temperature
$kT$~=~15.4~$\pm$~1.4 keV (Asaoka \& Koyama 1990).
While the thermal bremsstrahlung
yields an acceptable fit to the data, not only is the temperature
too high but the upper limit
on the equivalent width of the Fe-K line emission (less than 0.17~keV) is too
low to accommodate the expected value of emission from a plasma at cosmic
abundance in ionization equilibrium.
Although the properties of \g21\ suggest the presence of a pulsar, 
X-ray and radio imaging  show no
indication of a point source (Becker \& Szymkowiak 1981, Kaspi \ea \ 1996), 
and all the searches for pulsation at any wavelength have proved 
inconclusive.
\g21 \ was recently observed with the \chandra \ X-ray
observatory as a calibration
target. Slane \ea \ (2000) discovered a low-surface-brightness
 component extending to a radius of $\sim$2\am \ from
the central core, using
a 34 ksec exposure with ACIS-S and a 30 ksec exposure with HRC-I. 
The authors interpret this newly discovered 
component as the first evidence for the expanding ejecta
and blast wave formed in the
initial explosion.

In this study we have carried out a combined analysis of 
 \chandra, \rosat, and \asca \ datasets.
 We focus on the \chandra \ data
which provide the best angular resolution to date. 
We compile all publicly available observations by \chandra \ and present here
the imaging, detailed spectral and timing analysis of a net 
72~ksec exposure 
with ACIS-S and 76~ksec observation with the HRC. This allowed us to
examine in more detail the spectral variations across
the remnant and put a more stringent upper limit on the pulsed fraction from
the yet undetected pulsar. With the longer exposure with ACIS,
we detect the low-surface-brightness, extended component
(referred to as the `shell' in Slane \ea \ 2000)
out to a radius of 2\am.5, and 
rule out its thermal interpretation.
 We conclude that the spectrum of the entire remnant
is dominated by non-thermal emission.  The broadband images, the hardness
ratio map, and the spectral properties of the remnant indicate that
the entire X-ray emission could be due to synchrotron radiation from
high-energy electrons injected by the `hidden' pulsar in the form
of either a wind or jets interacting with the surrounding medium.
We discuss our interpretation in comparison with the high-resolution 
22.3 GHz observations and compare the parameters of \g21 plerion to
the well studied Crab and 3C~58 plerions.
While \g21\ bears similarities to plerions, it has unique
properties that makes it an intriguing object.
To date, it is the only plerion whose X-ray size is bigger than
in the radio.  Its study is important to shed light
on plerions of the `second kind', a subclass of plerions
having a spectral break at millimeter wavelengths 
(Woltjer et al. 1997).
The paper is organized as follows:
In the following section, we describe the observations. 
We describe our imaging analysis in \S~3, and the corresponding spectral 
analysis in \S~4. In \S~5, we present the results of our timing 
analysis. Finally, we discuss our results in the last section.

\section{Observations}

\subsection{\chandra}

Because of its power law spectrum,  small size and relative brightness
(its unabsorbed flux is $\sim$4.3$\times$10$^{-11}$ ~ergs~cm$^{\rm -2}~s^{\rm -1}$
between 0.4 and 2.0 keV),
\g21\ was selected as one of the calibration targets of the \chandra\ observatory
(Weisskopf,  O'dell, \& van Speybroeck 1996).
As such, the SNR was observed in many configurations chosen
by the instrument teams.
In the following analysis, we have made use of these different configurations
to probe different aspects of the remnant.
Because of the CTI problems\footnote{
http://cxc.harvard.edu/cal/Links/Acis/}
with the front-illuminated  chips, we have not
used any data from the ACIS-I, even for imaging purposes (the spatial
resolution of about 1\as \ of the damaged CCDs is still intact) because
the CTI problems make it hard to perform accurate energy cuts on the data. \\   

For the spectral analysis, we used six observations (ObsID~159, 1230,
1433, 1716, 1717, and 1718) corresponding
to a total exposure time of 71.6~ks and obtained with  
 ACIS-S3 (back illuminated chip) for which the most accurate
calibration exists at this point.
High-resolution imaging and timing analysis were done 
using both the HRC-I and HRC-S and 
their unparalleled angular resolution.

\subsection{\rosat \ and \asca}

\g21\ was observed with the \rosat\ observatory using both the
Position Sensitive Proportional
Counter (PSPC) and the High-Resolution Imager (HRI).
Both observations are public\footnote{Data are available through the
High Energy Astrophysics Science Archive Research Center (HEASARC) at
 http://heasarc.gsfc.nasa.gov/},
and although they are quite short (7.2~ks for the PSPC and
4.7~ks for the HRI), we use the PSPC data in the following analysis
to better constrain the value of the column density and the HRI data to
compare with the images extracted from the \chandra\ datasets.\\
\g21\ was also observed by both the Solid-State Imaging Spectrometer (SIS) and
the Gas Imaging Spectrometer (GIS) on-board \asca \ and the data are also
available through the HEASARC database. The exposure time
is 28 ksec with the GIS and 24 ksec with the SIS.  
Because of the relatively poor angular resolution (3\am) of the \asca\
X-ray telescopes,
none of the fine spatial structures of \g21\ can be studied with neither the
GIS  nor the SIS.  \asca\ advantages reside in the large number of counts collected
and the GIS
timing capabilities (a timing resolution of 6.1$\times$10$^{-5}$s for the
high-bit rate mode in which the data were taken).
We have used the spectral analysis of the
\asca\ datasets as a measure of the ``average'' spectrum from the remnant
to complement that extracted from the \rosat\ PSPC, and in combination with
that of \chandra, which is the focus of this paper.

\section{Imaging}

We have screened the ACIS-S data
by running {\it fselect} on the primary events files.
Only events with standard ASCA grades 02346 are retained.
 We rejected columns which lie on the edges of read-out
node areas and where event selection will not operate correctly since the
charge cloud is split between 2 different read-out nodes.
We rejected the hot columns and pixels that are listed on the
Chandra X-ray Center web site\footnote{http://asc.harvard.edu/cal/Links/Acis/}.
We have extracted images of \g21\ in different energy bands. 
 The images are centered at
RA (J2000) = 18$^{\rm h}$33$^{\rm m}$33$^{\rm s}\!.$50,
and DEC (J2000) = $-$10$^\circ$34$^\prime$6$^{\prime\prime}\!.$6.
In Fig.~1, we show the soft (0.1--6~keV) and hard (6--10~keV) band images, 
smoothed with a Gaussian with $\sigma$~=~3\as.
While the extended component is evident in the soft band image, 
only the inner core is left in the hard band.
The inner core has an angular radius of 40\as, similar
to the radio plerion. We refer to this component
as the inner core throughout the paper.
The extended component extends
out to a radius of 150\as, slightly bigger
than previously reported (120\as, Slane et al.
2000), and confirmed by \xmm \ (Warwick et al. 2000).
We refer to this component (50\as--150\as) as the low-surface-brightness
extended component throughout the paper.

In Fig.~2, we show the hardness ratio map using
the ratio: (2.4--10) keV over (0.5--2.4 keV).
The image is smoothed with a Gaussian with $\sigma$~=~3\as.
 It is evident from the image that the central
core is harder than the extended component.
Moreover, the extended component
is harder in the northern quadrant, with  brighter knots and filamentary
structures extending from the core out to the outer part of the SNR.
This result is consistent with the detailed spectral analysis described below
(\S~4.1).
 
In Fig.~3, we show the ACIS-S3 image in the 0.5--2.4 keV band
with the \rosat \ HRI contours overlayed. We smooth the \chandra \
image with a Gaussian with $\sigma$ = 6\as, to match the
angular resolution of the HRI.
The HRI image reveals the $\sim$40\as \ radius inner
core detected in the radio.
The extended component is detected with ACIS
out to a radius of 150\as.
The point-like source south-west of the plerion coincides with
an emission line star, SS~397.

\g21\ was observed five times with the
HRC-I for a total of more than 100~ks. 
Unfortunately, two of those observations were unusable because of
processing or reconstruction problems. 
We have used the three remaining observations and produced an image 
using a mosaic program which takes into account the difference
in pointing. 
The background is subtracted using a script, 
{\it{screen}}$_{-}${\it{hrc}}, made available by the HRC
Calibration team\footnote{http://asc.harvard.edu/cal/Links/Hrc/CIP/filter.html}. 
This procedure flags events with a non-X-ray charge 
cloud distribution and removes about 30\%--40\% 
of the background events while removing only a few percent
of real events.
We use this script and find that 33\% of events are cut, 
leaving a total of about 3 $\times$ 10$^6$ events.
We do not correct for exposure.
The result is shown in Fig.~4.  
The image has been convolved with an elliptical top hat filter whose
size varies to have a minimum of 16 counts under the filter. The pixel
size is 0\as.15. We find that 
the center of the HRC emission is located at
RA~(J2000) =
18$^{\rm h}$33$^{\rm m}$33$^{\rm s}\!.$51 ($\pm$0.01),
Dec~(J2000) = 
$-$10$^\circ$34$^\prime$8$^{\prime\prime}\!.$47 ($\pm$0.15).   
We did not measure the spectral difference across the 
different parts of the HRC-I emission because of the lack of spectral response
in the HRC.

In Fig.~5, we show the background-subtracted 
radial profile of the plerion obtained with
HRC and ACIS. We choose the observations with minimal offsets
from the telescope axis, for which a 0\as.5-radius aperture encircles
90\% of the point-source counts. We note that  
while the ACIS profile is shown out to a radius of $\sim$ 100\as,
the HRC profile extends to a radius of 40\as \ only due to
the high background at larger radii. 
In the inset, we show the HRC profile with 0\as.25 bins along with
the expected point-spread function (dashed line).
This shows that the brightest inner core is not a point source
and that the brightness drops sharply beyond a radius of $\sim$~0\as.5.
We use this profile later (\S~6.2) to infer
the luminosity of the putative pulsar in \g21.

\section{Spectral analysis}

\subsection{ACIS-S}

In the following, we present the spectral
analysis of the SNR and examine the spectral
variations on the arcsecond scale.
 For the inner 40\as \ core, we extract spectra from concentric rings
of width 5\as \ and centered at
RA~(J2000) = 18$^{\rm h}$33$^{\rm m}$33$^{\rm s}\!.$5,  
Dec~(J2000) = $-$10$^\circ$34$^\prime$6$^{\prime\prime}\!.$6.
We also extract the spectrum of the brightest 1\as \ radius central 
part of the image.
For the extended component, we extract a spectrum
of a ring extending from 50\as \ to
150\as---the latter defines the outer boundary detected with \chandra.
 
The background is extracted from the same detector region on the S3 chip
using an observation of \g21 obtained with the four ACIS-I chips and
  two ACIS-S chips activated.
This method allows one to account for both the chip and sky background.
The background-subtracted count rates in the 0.5--10 keV energy range
are 2.544 $\pm$ 0.008 \cps
and 0.506 $\pm$ 0.005 \cps from the inner core and the extended component,
respectively.
We processed the data as described in the 
imaging section.
We used CIAO v1.1 $\it{mkarf}$ tool in order to generate 
the corresponding ancillary response files.
The innermost 1\as-radius core has a very hard spectrum and is best described
by a power law model with
$\Gamma$ = 1.4 $\pm$ 0.1 ($N_H$ = 2.1 $\pm$ 0.1
 $\times$ 10$^{22}$~cm$^{-2}$), and
an unabsorbed flux in the 0.5--10~keV of 3.5~$\times$~10$^{-12}$~\ergpcps.
We subsequently fit all the spectra simultaneously with a power law
and find that the power law index steepens with increasing radius.
When allowing $N_H$ to vary, we get an acceptable fit
($\chi^2_{\nu}$ = 0.68, $\nu$ = 4451) with the photon index
increasing from 1.62 $\pm$ 0.05 at the inner
5\as \ circle ($N_H$ = 2.37 $\pm$ 0.08 $\times$ 10$^{22}$~cm$^{-2}$)
to 2.36 $\pm$ 0.07 at the outer ring
 ($N_H$ = 1.83 $\pm$ 0.07 $\times$ 10$^{22}$~cm$^{-2}$).
The variation of $N_H$ across the plerion could be an artifact
of the spectral fitting. 
When tying $N_H$, we get acceptable fits for all the spectra
($\chi^2_{\nu}$ = 0.71; $\nu$ = 4459), with
$N_H$= 2.24 $\pm$ 0.03 $\times$ 10$^{22}$~cm$^{-2}$
(a value consistent with that found by Warwick et al. 2000),
 and $\Gamma$
varying from 1.53 $\pm$ 0.03  (inner 5\as \ core)
to 2.73 $\pm$ 0.04 (the extended component).
All errors throughout the paper are at the 90\% confidence level.
We shall accept this
value as the $N_H$ for the SNR, with an uncertainty we shall determine
by the range of values we find for fits to other datasets below.
In Fig.~6, we plot the corresponding photon index as a function of radius
(with $N_H$ = 2.24 $\times$ 10$^{22}$ cm$^{-2}$).
The steepening of the photon index with increasing radius has been
noted by Slane \ea \ (2000) when fitting the inner 40\as \ core.
Slane \ea \ (2000) did not however rule out a thermal model for
the extended component, and interpret that component 
as the SNR shell or shocked ejecta. 
With this longer \chandra \ observation, we find that thermal models
are unlikely, as discussed further below.

In Table~1, we summarize the result of fitting
the extended component with thermal and non-thermal models,
in the 0.6-8~keV energy range.
A power law model gives the best fit;
with the following parameters:
$N_H$ = 1.83$^{+0.07}_{-0.06}$~$\times$10$^{22}$~cm$^{-2}$, 
$\Gamma$~=~2.36$^{+0.07}_{-0.06}$, and $\chi^2_{\nu}$~=~1.0
($\nu$=658, see Table~1).  
When freezing $N_H$ to 2.24~$\times~$10$^{22}$~cm$^{-2}$
(above paragraph),
we get $\Gamma$ = 2.73~$\pm$~0.04 and $\chi^2_{\nu}$~=~1.0 
($\nu$ = 659).
When fitting the extended component with thermal models,
we use the new models invoked in XSPEC version 11.0.1
\footnote{http://heasarc.gsfc.nasa.gov/lheasoft/xanadu/xspec/}.
We can reject a  CEI model (which describes a plasma in ionization equilibrium)
with solar abundances.
 In Fig.~7, we show the spectrum of the extended component fitted with a
power law (left) and $\it{APEC}$ model (right). The latter model 
is the new code of thermal plasma
including updated atomic lines
\footnote{http://hea-www.harvard.edu/APEC}.
It is clear that the $\it{APEC}$
model over-estimates the data in the Fe-K line region. A very low Fe
abundance is required to get an acceptable fit, with the
following parameters:  $N_H$=1.46~$\times$~10$^{22}$ cm$^{-2}$, $kT$~=~4.1
keV, Fe~=~Ni~=0 ($\leq$ 0.16, 2$\sigma$),
 $\chi^2_{\nu}$~=~0.96 ($\nu$~=~657). While the fit is formally acceptable,
the derived $N_H$ is inconsistent with that of the SNR.
We have subsequently
fitted the extended component with non-equilibrium ionization (NEI) models,
 which are more appropriate for modeling SNRs whose age is 
smaller than the time
required to reach ionization equilibrium.  We use the
\shock model (Borkowski, Lyerly, and Reynolds 2000), which
comprises a superposition of components of different ionization ages
appropriate for a plane-parallel shock. This model is characterized by
the constant electron temperature, $T_e$, and the shock ionization age,
$n_e t$ (where $n_e$ is the
postshock electron density, and  $t$ is the age of
the shock).
This model yields an adequate fit, but with an unusually low ionization
time-scale and again a very low value of $N_H$. The parameters are: 
$N_H$~=~1.53~(1.48--1.57)~$\times$10$^{22}$~cm$^{-2}$,
$kT_s$~=~3.50~(3.28--3.74)~keV, $n_et$~=~1.1~(0.8--1.3)~$\times$10$^9$
cm$^{-3}$~s, and $\chi^2_{\nu}$~=~0.86 ($\nu$=657).  
 We summarize these results in Table~1.  
When fitting with the \nei model, which is characterized by 
a constant-temperature and a single-ionization timescale, 
we find a similar result to the \shock model.
Varying the abundances of Mg, Si, S, Fe, and Ni (tied to Fe),
the \shock and \nei models yield a much higher $n_0~t$ ($\sim$10$^{12}$ cm$^{-3}$
s), but require unusually low abundances.  These models mimic
the CEI model with low abundances.  
We conclude that while thermal NEI
models give adequate fits, they either require an unusually low $n_et$
(with solar abundances),
or alternatively a large $n_et$ with unusually low metal abundances.
Furthermore, the derived $N_H$ from any thermal model is inconsistent with
that derived for the inner core. 
To account for a change of $\delta(N_H)$~$\geq$~0.5$\times$10$^{22}$~cm$^{-2}$
$\sim$ $\int n_e \ dl$,
an unrealistic local electron density of $n_e$ $\geq$ 10$^{3}$~cm$^{-3}$
is needed for a path $l$ $\sim$ 2.4~pc (the extent of the extended component).

We subsequently examined the spectral variations across the SNR.
For that purpose, we extracted spectra from the brightest knots 2\am \ north
of the core, the north-eastern edge of the SNR, and an inner circle
 south-east of the inner core.
The selected regions are shown in Fig.~8 (North is up, East is to the
left), and
 their corresponding spectra (in the 0.5 -- 8.0 keV) are shown in Fig.~9.
There is no evidence of line emission.  
 All the spectra are
well fitted with a power law model with $N_H$ = 1.9 $\times$10$^{22}$~cm$^{-2}$
(tied for the three spectra),
$\Gamma$= 2.14, 2.19, and 2.5 for the northern knots, the north-eastern edge,
and the south-eastern region,
respectively ($\chi^2_{\nu}$ = 0.50, $\nu$~=~416).
The spectrum of the northern quadrant seems to
have a harder spectrum that the rest of the diffuse emission. 
This result is in agreement with the hardness ratio map shown in Fig.~2.

\subsection{\rosat \ PSPC}
The source spectrum of the inner core is extracted from a circle
of radius $R$~=~40\as.
This defines the extent of the plerion seen with \rosat \ (Fig.~3).
When fitting the PSPC spectrum in the 0.1--2.4~keV with a power law model,
we get $N_H$ = 1.95 (1.35--2.74)$\times$10$^{22}$~cm$^{-2}$,
$\Gamma$~=~1.95 (0.5--3.8), and $\chi^2_{\nu}$~=~0.95 ($\nu$ = 75).
The errorbars are large due to the poor statistics and the narrow
energy band of \rosat. To better constrain the parameters,
we fit the \rosat \ PSPC (0.1--2.4~keV) spectrum
simultaneously with the ACIS-S (0.5--10~keV) spectrum extracted
from the 40\as-radius core.
The power law model yields: $N_H$ = 2.23 (2.22--2.76)$\times$10$^{22}$
cm$^{-2}$, $\Gamma$ = 1.85 (1.83--1.87), 
and $\chi^2_{\nu}$~=~1.1 ($\nu$~=~889).
The total observed flux from the inner core
is 5$\times$10$^{-11}$ \ergpcps,
which translates to an unabsorbed luminosity $L_X$ =
2.8$\times$10$^{35}$ erg~s$^{\rm -1}$ (at a distance of 5~kpc). 
The data and fitted model are shown in Fig.~10.

\subsection{Comparison with \asca \ archival data}

We here present the analysis of the \asca \ archival data in order 
to compare \asca \ and  \chandra\  spectra. While \asca \ has a much poorer 
angular resolution than \chandra, its spectral resolution is comparable
 to that of the ACIS-S3 chip. \asca \ has the advantage
over \exo \ and \gin \ 
in better determining the average photon index of the plerion.
This was noted by Gallant \& Tuffs (1998) who found that 
extrapolating from X-rays to the 
infrared wavelengths using the photon index
 determined with \exo \ and \gin \
yields fluxes lying well below the measured ISOCAM values.

We have extracted the \asca\ spectra from a circular region of radius
$R$~=~3\am, centered at the maximum of emission
 (RA (J2000) = 18$^h$ 33$^m$ 33$^s$.5, 
and Dec (J2000) = $-$10$^o$ 34$^{\prime}$ 3$^{\prime\prime}$.1).
We chose a region of 3\am \ radius since the source is 7\am.25 
\ offset from the telescope axis,
 and a larger extraction radius would 
include events from the edge of the detectors.
Since \g21\ is located
in the Galactic bulge region, we performed the background subtraction
using a region from the same field of view.  This has the advantage of
minimizing the contamination of the source spectra by the high-energy
emission from the Galactic ridge.
                                                   
We combine the spectra  from the SIS and GIS detectors, and fit them
simultaneously, by introducing a relative normalization between the SIS
and the GIS spectra.  A power law model yields an adequate fit with an
interstellar absorption
 $N_H$ = 2.06 (2.0--2.12) $\times$10$^{22}$ cm$^{-2}$, a photon index
$\Gamma$ = 1.89 (1.85--1.93), and $\chi^2_{\nu}$~=~0.98 
($\nu$=755).  The observed flux in the 0.5--10 keV
range is 7.6$\times$10$^{-11}$ \ergpcps,
which translates to an unabsorbed luminosity $L_X$ (0.5--10 keV) =
4.3$\times$10$^{35}$ erg~s$^{\rm -1}$ (at a distance of 5~kpc).
The derived photon index of 1.9 is steeper 
than that derived with \exo \ (Davelaar et al. 1986),
and is closer to the value of 2.0 required by Gallant and Tuffs
(1998) to account for the measured infrared fluxes. The 
small discrepancy might be due to the contamination of
the ISOCAM fluxes by emission lines or due to
subtraction of stellar contribution (Y. Gallant, private communication).

\section{Timing}

To search for pulsations in the \asca \ data, we extract the 
events from an aperture of radius 3\am \ in the GIS images.
The GIS data were obtained in the high-bit rate mode, 
with a photon time-of-arrival (TOA) resolution of 0.061
milliseconds.  We perform a power spectral density analysis on the
barycenter corrected photon arrival times using the hard band (2--10
keV).  No significant pulsations were found.
The failure to detect
pulsations could be partly attributed to the contamination of the
extracted events by the plerion.
The analysis of the \chandra \ ACIS-S3
image indicates that only
 $\sim$ 20\% of the counts
collected in the $3'$ \asca\ GIS PSF
originate from the
central 5\as \ radius circular region  (in the 2-10 keV band).

We have also searched for pulsations in the \chandra\/ HRC data.
Slane \ea \ (2000) analyzed one HRC-I set (ObsID {\it 1406}) and
obtained an upper limit of 40\%
on the pulsed fraction of the central source.
By the time of our analysis, there were eleven HRC observations in the
public archive.
We use only five of them (Table~2):
2 sets acquired with the HRC-S and 3 sets with HRC-I.
The rest of the HRC observations were observed with large $\pm$Y offsets
\footnote{http://asc.harvard.edu/udocs/docs/POG/MPOG/index.html}, 
that led to considerable spread of the source profile, or contained
serious processing errors (ObsID {\it 1298}). 
Various parameters of the analyzed sets (including the time spans
$T_{\rm span}$ and the effective exposures $T_{\rm exp}$)
are presented in Table~2.
All the data were obtained with non-optimal focus positions
(see parameter SIM\_X in Table~2), 
which led to broadening of the core.
For each of the 5 sets, a short period of time at the start of the
exposure was excluded from consideration because of
an unstable count rate. The total time span for all the observations
(Sep 4-- Feb 15) amounted to 14,328~ks,
with 75.731~ks total effective exposure time.

The HRC-I data were filtered making use of the HRC screening algorithm%
\footnote{http://asc.harvard.edu/cal/Links/Hrc/CIP/filter.html}.
The TOA's were converted to barycentric dynamic times (TDB)
with the {\it aXbary} task of CIAO.
We extracted the 
events for the timing analysis from the apertures 
of 15-pixel ($\approx 2''$) radius centered at the core position,
the numbers of counts are given in Table~2.
The event count rate in set {\it 143\/} is lower than in the other sets
because its higher background
count rate saturated telemetry.

Because of a wiring error in the HRC 
detector\footnote{http://asc.harvard.edu/udocs/hrc/timing.html}, 
the TOA
ascribed to an event in a data file is actually the TOA of the following
event, which may be a photon event or, more likely, a background particle
event not telemetered to the ground.
Therefore, a typical TOA error is equal to the average time between
the successive events in the HRC detector, including the particle events.
Making use of the Level~0 supplementary data products, which contain
the total (photon + particle) count rates for the whole HRC detector,
we estimated
the mean errors of about 5 ms for all the sets except for {\it 143},
for which the error is about 1 ms. In principle, the errors could be
reduced by additional filtering of the extracted events, retaining
only events such that the time to the next recorded event is less than
a given value. This filtering, however, strongly reduces the number
of events available for the analysis, so it is not applicable in our case
 due to the small number of photons.
Therefore, the timing errors force us to choose a minimum period 
 considerably longer than
5 ms (maximum frequency smaller than 200 Hz).
We chose the frequency range $0.01 < f < 50$ Hz.

The presence of several data sets allows us to use the ``correlated periodicity
search'' (Pavlov, Zavlin \& Tr\"umper 1999). For each of the data sets, we
use the $Z^2_1$ (Rayleigh) test (Buccheri et al. 1983). The function $Z_1^2(f)$
has numerous peaks, with characteristic widths 
$\delta f \approx T_{\rm span}^{-1}$; one of these peaks may correspond to
the pulsar frequency, the others are caused by statistical noise.
If there is the pulsar peak at a frequency $f_0$
 in one of the data sets, it should also be
seen in each of the other sets, at frequencies $f_0+\dot{f}_0\Delta T$,
where $\dot{f}_0$ is the pulsar frequency derivative,
and $\Delta T$ is the time between the corresponding observations.
To identify the true pulsar peak (i.e., to find $f_0$ and $\dot{f}_0$), 
we choose a reference data set
and examine highest peaks in this set for the presence of counterpart
peaks in the other sets, for a reasonable range of $\dot{f}$.
If no counterpart peaks are found in at least one of the data sets,
we have to conclude that the peak
under examination is due to statistical noise. If counterpart peaks
are found in all the sets, they must satisfy some statistical criteria
to be considered viable candidates for the pulsar peaks (see 
Koptsevich and Pavlov 2001
for a description of the algorithm). 

For the frequency derivative
range, we choose $-1\times 10^{-9} < \dot{f} < 0$ Hz~s$^{-1}$, which includes
the $\dot{f}$ values for all the known rotation-powered pulsars.
All the sets, except for {\it 1406\/}, meet the requirement 
$|\dot{f}|_{\rm max} T_{\rm span}^2/2 \ll 1$,
which allows us to consider $Z^2_1$ as independent of $\dot{f}$
for each of these sets.
To make the computations uniform, we divide set {\it 1406\/} in two 
successive subsets, {\it 1406$^a$\/} and {\it 1406$^b$\/}
(see Table 2), which satisfy this requirement.

We applied this approach to the six data sets and found no statistically
significant periodicity. The series of highest peaks found
in all the six sets corresponds
to ephemeris parameters $f=47.7154362408$ Hz and
$\dot{f}=-6.60\times10^{-10}$ Hz~s$^{-1}$ (epoch
51425.25047454 MJD TDB). Combining the 6 sets in one,
we find a $Z^2_1$ value of 36.6 for these $f$ and $\dot{f}$.
The probability to find such a peak
is close to unity, so that the result is statistically insignificant.
The corresponding pulsed fraction for a sinusoidal signal, $f_p
=(2 Z_1^2/N)^{1/2} = 16\%$ (where $N=2881$ is the total number of events),
coincides with that measured from the extracted light curve
($15\%\pm 5\%$) and
can be considered as an upper limit. 
This is 
considerably smaller than the limit
derived by Slane \ea \ (2000).

\section{Discussion}

We have presented archival X-ray observations of \g21 with \chandra,
\rosat, and \asca.
 The spectrum of \g21 is non-thermal and best fitted with
a power law model, whose index steepens by $\Delta$$\Gamma$$\sim$1
from the central part of the remnant to
the outer component (at $\sim$150\as).
The total unabsorbed flux from the 5 arcminute-diameter
SNR is $F_X$ (0.5 -- 10 keV) = 1.1 $\times$
10$^{\rm -10}$~erg~cm$^{\rm -2}$~s$^{\rm -1}$ with an 85\%
contribution from the 40\as \ radius core. At a distance of
5 kpc, this translates to a total luminosity
$L_X$ (0.5 -- 10 keV) = 3.3 $\times$ 10$^{35}$
erg~s$^{\rm -1}$.
The low-surface-brightness extended component discovered with \chandra \ does not have the 
limb-brightened morphology expected from a SNR blast wave or shocked
ejecta. It has rather a filled morphology, with knots and filaments 
most evident in the northern quadrant.
There is no significant evidence of thermal emission lines from this component.
Based on the morphology and the
spectrum of this component, we conclude that it is an extension of the 
inner core, and that the filaments could be plerionic `wisps'
as seen in Crab-like plerions.
 We note that \xmm \ has recently observed \g21 \ (Warwick et al 2000), 
and the results agree with our findings. 
In this paper, we have investigated in more detail the
spectral variations across the remnant on smaller scales, 
and determined a more stringent upper limit on the
pulsed fraction.
We have also presented for the first time the results from the
 \rosat \ and \asca \ observations,
as a consistency check for the overall spectral analysis.
We show that only the non-thermal model fits to
the extended component (with \chandra) yield a column density 
consistent with that found for the core, whereas the thermal models
(CEI or NEI) yield a much lower $N_H$ value.

In the following, we use our spectral and timing results to determine the
parameters of the plerion, and a `hidden' pulsar.
By comparing the \chandra
\ ACIS image with the radio image, 
we conclude that the morphology of \g21 \ could
be associated with a pulsar wind and possibly jets interacting with the surroundings.
The failure to detect pulsations could be due
to a geometrical effect.
We estimate the parameters of the putative pulsar in \g21 plerion,
and compare them with
 those derived for the well-studied Crab and 3C~58 nebulae.
Finally, the absence of thermal
emission in \g21 could be due to a low density medium
or a cavity in which the supernova explosion occurred. 

\subsection{Distance}

To estimate the distance to \g21, we
regard the $N_H$ values found from the ACIS fits, and those from the ASCA and
ROSAT fits, to describe a reasonable range of possible $N_H$ values,
(1.8 -- 2.3) $\times$ 10$^{22}$ cm$^{-3}$.  This range encompasses the values from all the
global power-law fits, and excludes the considerably different values
required for thermal fits.
The extinction per unit distance in the direction of \g21
can be estimated from the contour diagrams given by Lucke (1978):
$\langle E_{B-V}\rangle /D \sim 0.8$~mag~kpc$^{-1}$.
Using the relation 
$\langle N_H/A_V \rangle = 1.79\times 10^{21}$~cm$^{-2}$~mag$^{-1}$,
which translates into
$\langle N_H/E_{B-V}\rangle = 5.55\times 10^{21}$~cm$^{-2}$~mag$^{-1}$
(Predehl \& Schmitt 1995),
we derive a distance of 4.1--5.2 kpc.
This range is consistent with the lower estimate on
$D$ obtained
from 21 cm absorption (Becker \& Szymkowiak 1981). 
In the following, we assume a distance of 5~kpc and 
express our results as functions of $D_5=D/(5~{\rm kpc})$.

\subsection{The putative pulsar}

The morphology and spectrum of the inner core indicate the
presence of a pulsar powering \g21. The failure to detect pulsations
could be due to a beaming effect. We note that
3C~58 is another Crab-like plerion whose properties hint
at the presence of a pulsar, yet 
no pulsar has been found there
(Torii \ea \ 2000). To estimate the 
spin-down energy
loss of the pulsar, $\dot{E}$, 
one can use an empirical relationship 
connecting $\dot{E}$ and the X-ray luminosity, $L_X$,
for pulsar-powered plerions.
 For instance, using the formula by Seward \& Wang (1988),
$\log L_X= 1.39 \log\dot{E} - 16.6$,
where $L_X$ is the luminosity of the pulsar plus plerion in
the (0.2--4)~keV band,
we get $\dot{E} = 2.7 \times 10^{37} D_5^{1.44}$~erg~s$^{\rm -1}$.
Using $L_X=3\times 10^{-11} \dot{E}^{1.22}$ erg~s$^{\rm -1}$
(Slane \ea \ 2000) and
a total luminosity (inner core+extended component)
of 3.3 $\times$ 10$^{35}$ $D_5^2$ erg~s$^{\rm -1}$ in the 0.5--10 keV band,
we estimate $\dot{E}$ = 5.5 $\times$ 10$^{37}$ $D_5^{1.64}$ erg~s$^{\rm -1}$.
Thus, a plausible estimate for the spin-down energy loss is 
$\dot{E}_{37}\equiv \dot{E}/(10^{37}~{\rm erg}~{\rm s}^{-1})\sim 3$--6.

We can now use this $\dot{E}$ and a plausible age of \g21
to estimate the parameters of
the putative pulsar powering \g21.
The age of \g21 is
rather uncertain --- various estimates range 
from 0.8 to 40 kyr
(Salter \ea \ 1989, F\"urst \ea \ 1988, Morsi \& Reich 1987).
%
The linear size of the X-ray plerion in \g21 is
$\sim 3.6 D_5$~pc diameter, slightly bigger than that in
the Crab and 3C~58 nebulae.
The ratio $L_X$/$\dot{E}$ for \g21 is 
$\sim 0.006$--0.011.
This is smaller than those 
for the 1,000-yr-old plerions around the Crab pulsar and
PSR 0540--69 (0.05), 
and it is comparable with $L_X/\dot{E} \approx 0.01$ for the plerion
associated with the 1,500-yr-old pulsar, PSR~1509-58
in the SNR MSH~15-52 (Chevalier 2000).
On the other hand, the known pulsars with $\dot{E}_{37} \sim 3$--6 have characteristic
ages, $\tau$, in a range of 1.5--10 kyr.
In the following analysis, we normalize the pulsar parameters to $\tau_{3} =\tau/(3~{\rm kyr})$.

Given $\dot{E}$ and $\tau$,
the period $P$ of a pulsar can be estimated as
\be
P=\left[\frac{4\pi^2I}{(n-1)\dot{E}\tau}\right]^{1/2} =
0.144 \left(\frac{2}{n-1}\right)^{1/2} \left(\frac{I_{45}}{\dot{E}_{37} \tau_3}
\right)^{1/2}~{\rm s}~.
\ee
where $n$ is the braking index ($\dot{P}=K P^{2-n}$),
$\tau = P/[(n-1)]\dot{P}]= t + \tau_0$,
$t$ is the true age, $\tau_0=P_0/[(n-1)\dot{P}_0]$, $P_0$ and $\dot{P}_0$
are the initial period and its derivative,
and $I = 10^{45} I_{45}$ g~cm$^2$ is the moment of inertia. 
In this model, $\tau \approx t$
if $t\gg \tau_0$, which is equivalent to $P\gg P_0$.
The corresponding period derivative is 
$\dot{P}=P/[(n-1)\tau] = 7.6\times 10^{-13} [2/(n-1)]^{1/2} I_{45}^{1/2}
\dot{E}_{37}^{-1/2}\tau_3^{-3/2}$.
Assuming $n=3$ (magneto-dipole braking), we can estimate the 
conventional magnetic field,
$B=3.2\times 10^{19} I_{45}^{1/2} R_{10}^{-3} (P\dot{P})^{1/2}=
3.2\times 10^{13} I_{45}^{3/2} \dot{E}_{37}^{-1/2} \tau_3^{-1} R_{10}^{-3}~{\rm G}~,$
where $R=10 R_{10}$ km is the neutron star radius.
The inferred magnetic field is comparable to those of very young radio
pulsars.

The X-ray luminosities, $L_{X}^{\rm (p)}$
 of the radio pulsars detected with \rosat\
(0.1--2.4 keV band) show strong correlation with the value of $\dot{E}$,
which allows us to estimate the X-ray luminosity of the putative pulsar
in \g21. Using the relationship derived by
\"Ogelman (1995),
$L_X^{\rm (p)}\sim 6.6\times 10^{26} (B_{12}/P^2)^{2.7}$~erg~s$^{\rm -1}$,
 which is equivalent to
$L_X^{\rm (p)}
 = 1.5\times 10^{-16} \dot{E}^{1.35}$~erg~s$^{\rm -1}$,
we obtain
$L_X^{\rm (p)} = 1.3 \times 10^{34} \dot{E}_{37}^{1.35}~{\rm erg~s}^{-1}$.
The formula suggested by Becker \& Tr\"umper (1997), 
$L_X^{\rm (p)} =  10^{-3}\dot{E}$,
gives approximately the same value,
$L_X^{\rm (p)} = 1\times 10^{34}\dot{E}_{37}$  erg~s$^{\rm -1}$.
We can now compare these values with an observed upper limit on the pulsar
flux.
To estimate a conservative upper limit, we measure
the flux in a 0\as.5-radius aperture placed at the brightest part of
the core (see Fig.~5). 
We normalize the PSF to match the number of counts in the central
0\as.5 bin. We then estimate the ratio of counts within
0\as.5 to counts within 1\as \ radius and scale their
corresponding fluxes accordingly.
Using the estimated flux of  $3.5\times 10^{-12}$ \ergpcps \
from the 1\as \ central core (\S~4.1),
we estimate an unabsorbed flux of 2.4$\times$10$^{-12}$\ergpcps \
from the point source.
This translates to an upper limit on the flux from a hidden pulsar 
of $1.1\times 10^{-12}$ \ergpcps \ in the 0.1--2.4 keV range
(assuming $\Gamma =1.5$).
If the pulsar's X-ray emission were isotropic,
the corresponding upper limit on the pulsar luminosity would be
$L_X^{\rm (p)} < 3.5\times 10^{33} D_5^2$ erg~s$^{\rm -1}$,
an order of magnitude lower than the above predictions based on the $\dot{E}$ estimates.
This indicates that the X-ray radiation of the pulsar is strongly beamed,
and the orientation of the pulsar's rotation and magnetic axes is such
that the rotating beam does not come close to the line of sight,
in agreement with the lack of observable pulsations.

\subsection{Pulsar Wind Model}

In the Kennel and Coroniti (1984; hereafter KC~84) model, a pulsar injects high-energy
particles in the form of a relativistic particle wind.
Confinement of the pulsar wind by its surroundings causes
the outflow to decelerate at a shock, forming a synchrotron nebula.
The radius of the shock, $R_s$, is given by equating the pressure of the
pulsar's wind, $\dot{E}$/($4\pi R_s^2 c$), 
with the pressure in the nebula, $P_n$.
Beyond this radius, a non-relativistic flow transports the plasma from
the shock region to the edge of the nebula. 
The size of the nebula, $R_n$ is related to the shock radius,
$R_s$ as: $R_n$/$R_s$ $\sim$ 1/$\sqrt\sigma$; where
$\sigma$ is the so-called pulsar wind magnetization parameter defined as
the ratio of the Poynting flux to the particles flux (KC~84).
Self consistent models for the Crab 
require that the wind is particle dominated,
i.e. $\sigma$ is small ($\leq$~10$^{-2}$; KC~84).
From the \rosat \ PSPC and \chandra \ ACIS-S spectra, we estimate 
an equipartition magnetic field, $B_{eq}$ = 1.8$\times$10$^{-4}$~G 
in the 40\as \ radius core.
This yields a pressure in the inner 40\as \ radius
core of $B^2/8\pi$~$\sim$~1.29$ \times$10$^{-9}$ erg~cm$^{\rm -3}$,
and therefore $R_s$ = 0.046~$\dot{E}_{37}^{0.5}$~pc.
Using the estimated $\dot{E_{37}}$ $\sim$ 3--6 (above subsection), we get
$R_s$ $\sim$ 0.08--0.11~pc;
which corresponds to an angular size of $\sim$ 3\as--5\as \ (at 5~kpc).
Using the nebular size $R_n$ $\sim$ 150\as, we estimate
the pulsar wind parameter, $\sigma$ $\sim$~
($R_s/R_n$)$^2$ $\sim$ (4--11)~$\times$~10$^{-4}$,
indicating a particle dominated wind---like the Crab 
and the other Crab-like nebulae.
In Table~3, we summarize the inferred parameters of the
putative pulsar in \g21, in comparison with the Crab pulsar
and the putative pulsar in 3C~58.
When fitting the spectra in circular rings of \g21, 
we have found that the photon index
steepens away from the center all the way out to a radius
of 150\as---a result expected from synchrotron losses.
Chevalier (2000) has developed a one-zone, shocked wind model, to model
the X-ray luminosity of pulsar nebulae.
This model differs from the model of KC~84
in that the synchrotron burn-off 
leads to a steepening of the frequency spectral break by 0.5,
whereas it could be greater than 0.5 in the KC~84 model.
For \g21, the photon index, $\Gamma$, varies from 1.5 (in the innermost
5\as \ core) to $\sim$ 2.7 (in the outer diffuse component);
i.e. the net change of $\sim$~1.2 is inconsistent 
with the Chevalier model.

\subsection{Correlation with the radio image}

The high-resolution 22.3~GHz radio image of \g21,
obtained with the NRO Millimeter Array, shows a diffuse centrally 
peaked component plus axisymmetric filaments
(F\"urst et al. 1988). 
The authors suggest that the symmetry in
the lobes is evidence of precessing jets from a hidden pulsar.
In order to compare the radio image with our X-ray image, we
scale the ACIS-S image  to that of the radio 
image (Fig.~7 in F\"urst \ea \ 1988) and show the resulting 
image in Fig.~11.
The left panel shows the 40\as \ radius inner core of \g21, binned
to the same scale, smoothed
with a Gaussian with $\sigma$~=~3\as, and with the contours overlayed. 
The right panel shows the radio image of the core with the jets model
proposed by F\"urst \ea \ (1988). 
It is clear that the X-ray trough in the north-west is strongly
correlated with that of the radio image. Our result therefore supports
 the proposed picture. 
It is interesting that the shock radius ($R_s$ $\sim$ 0.1~pc; \S~6.3) 
coincides with the axisymmetric
filaments seen in the radio (F\"urst \ea \ 1988).
Therefore, those structures could be due to the deceleration of
the pulsar wind (or jets) confined by the nebular pressure. 
  They also found that the polarization 
varied from the center to the outer boundary, from
almost zero to 20\%--30\%.
They conclude that this is explained by a change in the magnetic field
geometry, with little contribution from a radial field from the inner
regions because of geometrical effects.

\subsection{Where is the SNR shell?}
To date, we know $\sim$ 10~Crab-like SNRs which lack 
 a SNR~shell, which would be expected from the
shocked ambient matter or shocked ejecta.
The lack of SNR shells is still a puzzling issue, and
many deep searches for shells around the Crab and 3C~58
have failed to find them. It has been suggested
that the absence of a shell could be due to a propagation
in a low density medium (in which case the shell would be
very faint), or to a low efficiency in accelerating particles
and amplifying the magnetic field 
(Reynolds \& Aller 1985).
A deep search in the radio for \g21 \ shell is needed, since all
current radio observations have been performed to study
mainly the bright inner core.
Slane \ea \ (2000) suggested that the extended component 
is associated with the blast wave of the SN explosion,
 or from the shocked ejecta.
Here we rule out thermal CEI models.
Fitting with NEI models, which are more realistic to model 
thermal emission from young SNRs,
indicates unusually low abundances or unusually
low ionization time-scales.
Furthermore, the derived $N_H$ is 
inconsistent with that derived for the SNR.
We find no evidence of line emission in any part of the remnant,
leading us to conclude that the spectrum is dominated by non-thermal emission.

A growing number of SNRs show evidence of 
non-thermal emission from their shells
(e.g. SN~1006). The emission mechanism
has been attributed to synchrotron radiation
from high-energy particles accelerated at the SNR shock (Allen,
Gotthelf, \& Petre 1999).
Reynolds and Chevalier (1981) have noted that the non-thermal spectra
should not be described by a power law, but rather a power law 
with a roll-off at high energies.
If the non-thermal emission in the extended component of
\g21 is indeed due to
synchrotron radiation from highly energetic particles 
accelerated at the SNR shock,
then one would expect it to be fitted with the cut-off or escape models
recently included in XSPEC v11.0, which are more appropriate for modeling 
non-thermal emission from SNR shells (Reynolds \& Keohane 1999, Dyer \ea \ 2000).
These models require knowledge of the radio flux density
and the radio spectral index.
With the absence of a radio counterpart,
we fit the X-ray spectrum with the {\it sresc}
model leaving these parameters free.
 This model does not provide a good fit ($\chi^2_{\nu}$ = 1.36,
$\nu$ = 670).  This again argues against a shell interpretation for
the extended component.

The non-thermal emission in the extended component is most likely due to synchrotron emission
from particles accelerated and injected by the `hidden' pulsar.
Using the power law model parameters of the extended component (Table~1,
1st row), and assuming equipartition of particles and magnetic field, we
estimate an equipartition field of 1.8$\times$10$^{-4}$~Gauss.
The lifetime of the synchrotron emitting electrons is only 
60~$E_{\gamma,keV}^{-0.5}$~$B_{-4}^{-1.5}$ years; where $E_{\gamma,keV}$
and $B_{-4}$ are the photon's energy in units of
1~keV and the magnetic field in units of 10$^{-4}$~G, respectively.
The synchrotron lifetime is much smaller than the estimated age of the SNR---
a result also found in the Crab and Crab-like nebulae.
The electrons can propagate out to a radius of 3~pc if
they are traveling at a speed of 0.15~c~$E_{\gamma,keV}^{0.5}~$~$B_{-4}^{1.5}$.
The enhancement of brightness in the northwest quadrant could 
be due to the propagation
into a denser medium (as evident in the public IRAS images of \g21).
To get an upper limit on the density of the ambient medium,
we fit the extended component
with a power law plus a thermal model ($\it APEC$).
We freeze the power law model parameters to their best value (Table~1), and
determine an upper limit on the emission measure ($EM$) of the thermal component.
We get $EM$ $\leq$ 1.1$\times$10$^{-4}$ (at the 90\% confidence level).
Using $EM$ =  10$^{-14}$~$(4 \pi D^2)^{-1}$~$\int n_e n_H dV$, where
$n_e$ and $n_H$ are the  postshock electron and Hydrogen density, respectively,
and $V$ is the volume of the extended component,
we estimate $\left<{n_e n_H}\right>^{1/2}$ $\leq$ 
 0.07$~D_{5}^{-1/2} \ {\rm cm}^{-3}$.     
Assuming a strong shock with compression ratio 4, 
the ambient density, $n_0$, is $\leq$~0.02~$D_{5}^{-1/2} \ {\rm cm}^{-3}$.

\g21 \ is perhaps most similar to the plerionic SNR 3C~58, although
there are some differences.
Both plerions have a morphology hinting at the presence of
a pulsar, yet undetected.
They both have a low-frequency spectral break (at 50 GHz), 
and thus belong to the class of plerions of the `second
kind' (Salvati et al. 1998).
Woltjer et al. (1997) suggest that these plerions
are powered by short-lived pulsars with either
very small braking indices or pulsars that would undergo
 a phase change in the pulsar's energy output.
 Both are young and exhibit a bright inner core plus
a fainter extended component whose spectra are dominated by non-thermal
emission.
The extended component in 3C~58 shows however a small flux (6\% of the total)
of thermal nature, which was attributed to the expansion of the outer rim into
inner ejecta core (Bocchino et al. 2001). 
\g21 \ is also similar to the Crab nebula in
that it has a centrally-filled morphology
and a non-thermal spectrum.
The softening of the spectral index
away from the center in \g21 \ is 
similar to that in the Crab nebula (Willingale et al. 2001).
In spite of these similarities, \g21 \ remains a unique
plerion.
Its X-ray brightness profile (core+plateau) 
and its morphology are puzzling,
 and can not be explained by
diffusion models such as Becker (1992).
 The detection of axisymmetric filaments in the 
radio and X-rays hint at a complex magnetic field geometry
and/or jet structures (Warwick et al. 2000).
\par
The most unusual property of \g21 \ plerion
lies in the radio regime.
Plerions are usually bigger in size at lower energies, since the
high-energy photons have a shorter lifetime than the low-energy photons.
Therefore, one would expect to see a bigger plerion in the radio.   
This is the case for the Crab nebula.
In 3C~58, there is a close correspondance
between the radio morphology of the outer nebula 
and the X-ray map at soft energies (Bocchino et al. 2001).
For \g21, only the inner 40\as\ radius core has been detected in the radio.
Slane \ea \ (2000) have put an upper limit
 on the 1~GHz surface brightness of the extended component of
$\Sigma$~=~4$\times$10$^{-21}$~W~m$^{-2}$~
~Hz$^{\rm -1}$~sr$^{\rm -1}$. 
Gaensler (2000, private communication) 
has recently found a radio knot coincident
with the hard X-ray brightest knot/filament found in the northern
region. 
A deep observation of \g21 \ in the radio and at the short wavelengths
(currently conducted with the VLA)
will determine the spectral characteristics of this knot and will
 allow us to search for the radio counterpart of the extended component
and the missing SNR shell.  

\acknowledgments{
This research has made use of the HEASARC database and the NASA Astrophysics
Data System (ADS).
We are grateful to the Center for Academic Computing of the Pennsylvania State
University for the opportunity to use the LION-X cluster} %
\footnote{http://cac.psu.edu/beatnic/Cluster/Lionx/lionx.html}
for the period search.           
We thank Keith Arnaud (GSFC) for his help with XSPEC and CHANDRA
software, Glenn Allen (MIT) for his help with ACIS issues
at the early stages of writing this paper, Slava Zavlin (MPE)
for useful discussions, Bryan Gaensler (MIT) for pointing
out the finding of a radio knot in the extended component,
and Yves Gallant (CEA, Saclay)
for his input on the infrared emission from \g21.
We also thank the referee for careful reading and useful comments. 
S.S.H. acknowledges support by the Natural Sciences and Engineering
Research Council (NSERC) of Canada.

\clearpage

\begin{figure}
\epsscale{1.2}
\plottwo{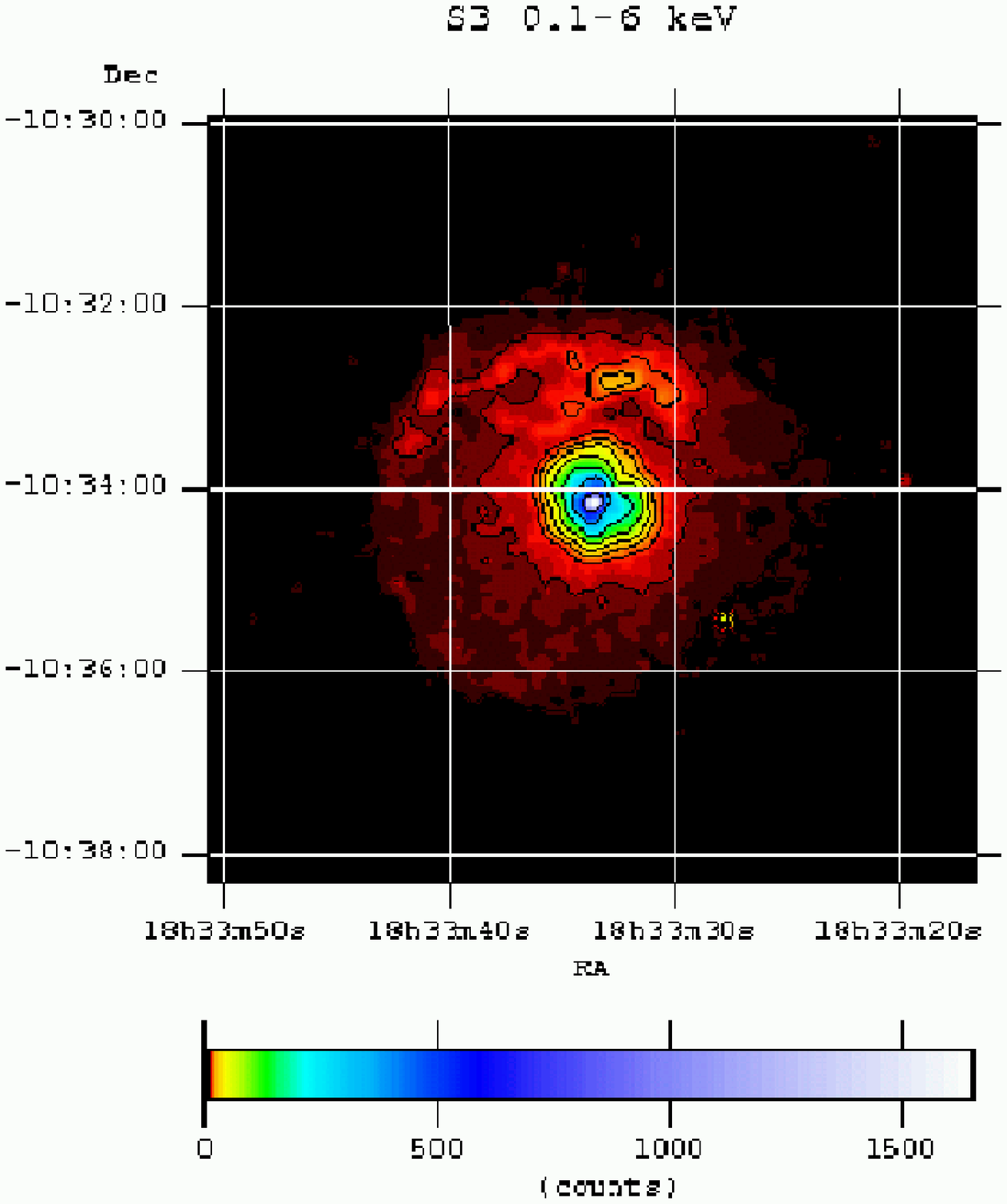}{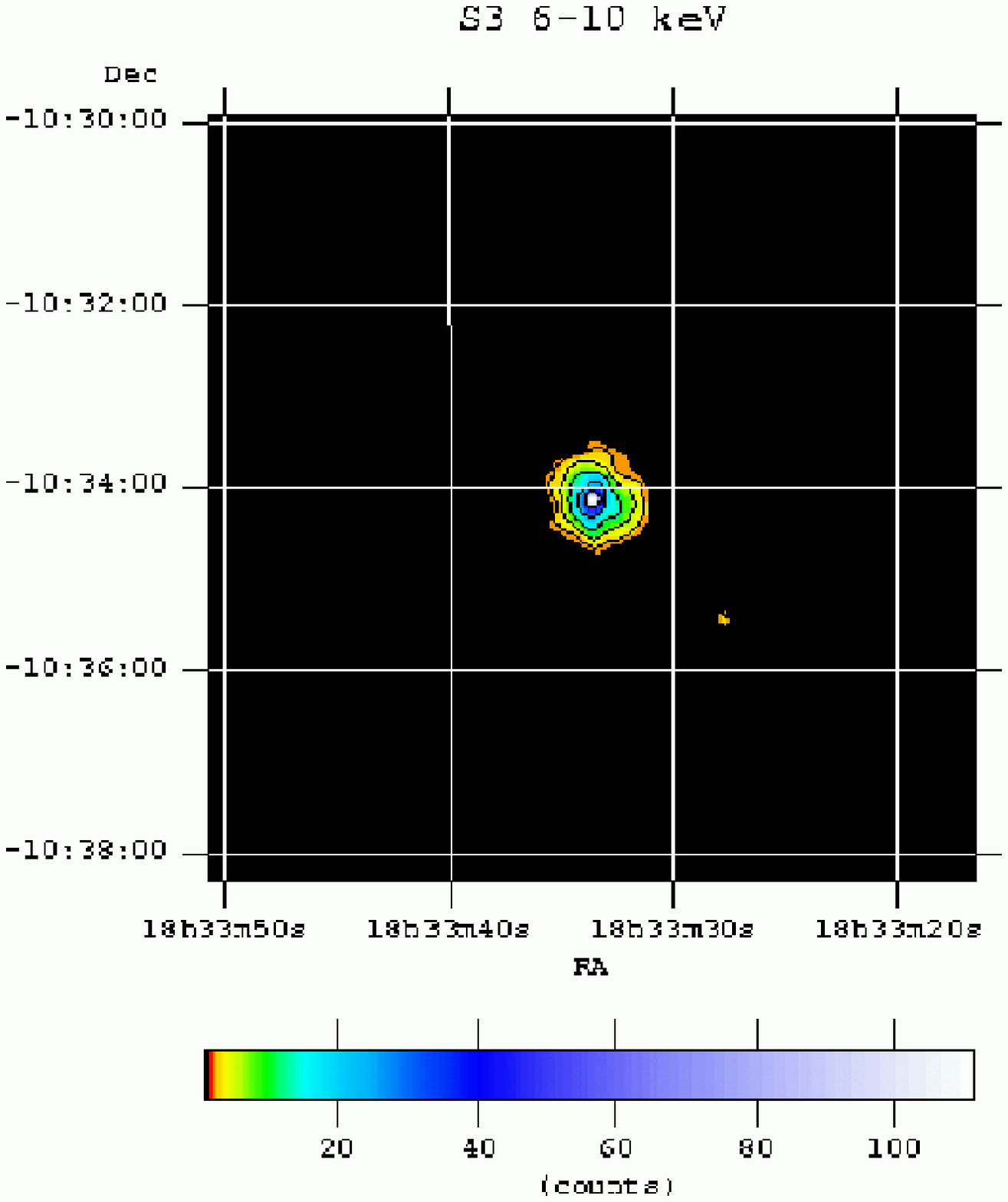} 
\caption{The \chandra \ ACIS-S images of G21.5-0.9 at
0.1--6.0 keV (top) and 6.0--10.0 keV (bottom).  The total
exposure time is 71.6 ksec. The images are smoothed
with a Gaussian with $\sigma$ = 3\as. Coordinates are J2000.} 
\epsscale{1.0}
\end{figure}

\clearpage

\begin{figure}
\epsscale{0.75}
\plotone{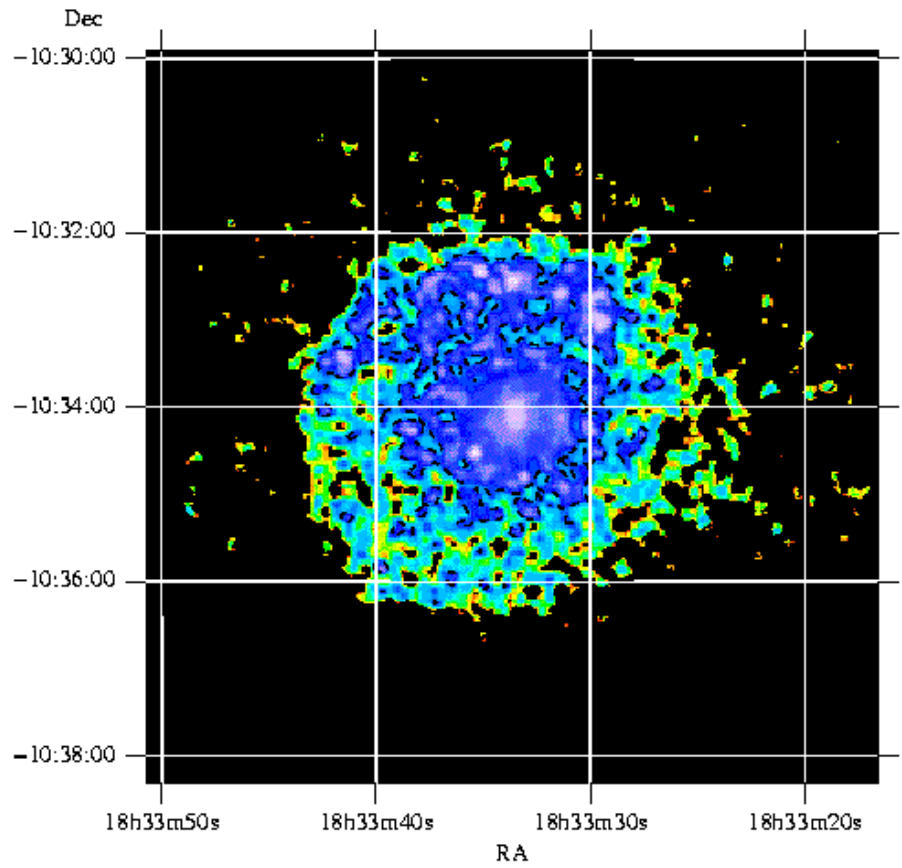}
\epsscale{1.0}
\caption{Hardness ratio map of G21.5$-$0.9 observed with
\chandra \ ACIS-S. 
The ratio is 2.4--10 keV over 0.5--2.4 keV. The image
is smoothed with a Gaussian with $\sigma$~=~3\as.
}
\end{figure}

\clearpage

\begin{figure}[tb]
\epsscale{0.75}
\plotone{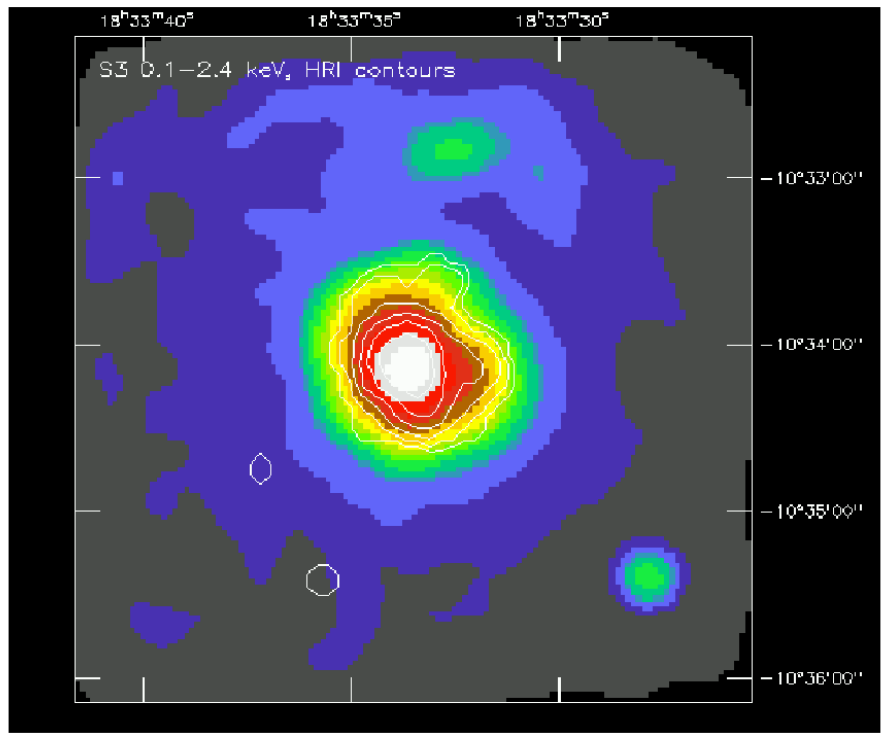}
\caption{
The ROSAT HRI contours of G21.5-0.9
showing the plerion ($R$$\sim$40\as)
overlayed on the soft band image
obtained with ACIS-S3. The point-like source
southwest of the plerion coincides with an emission-line star, SS~397.
Both the HRI and ACIS-S3 images are
smoothed with a Gaussian with $\sigma$~=~6\as \ and contours
are displayed in a log scale.}
\end{figure}    

\clearpage

\begin{figure}[tb]
\epsscale{0.6}
\plotone{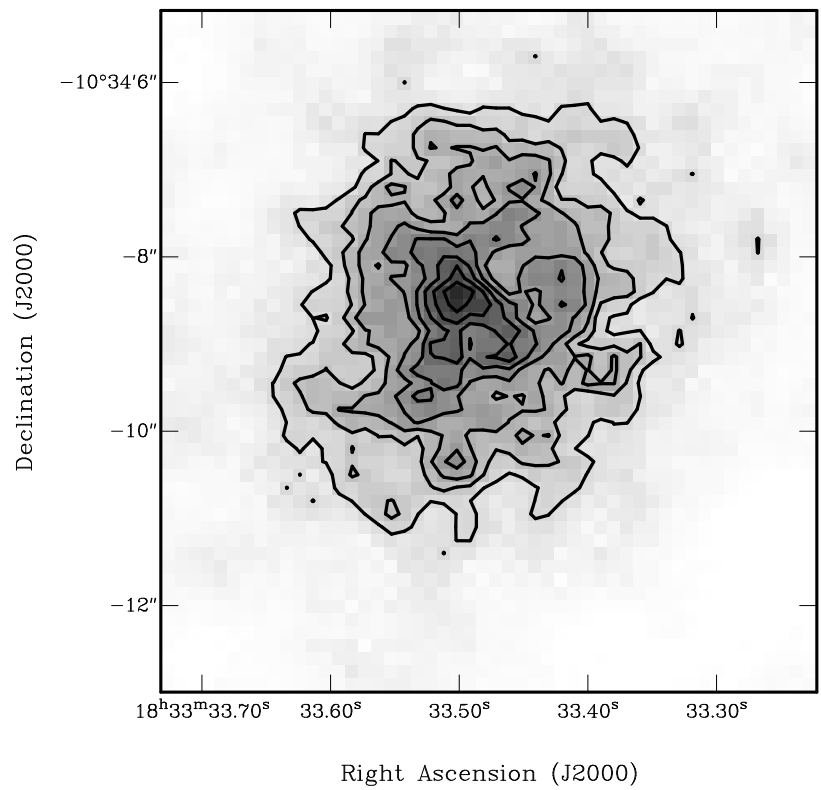}
\caption{The HRC-I image of the inner core of \g21.
See \S 3.3 for details. }
\end{figure}

\clearpage 

\begin{figure}
\epsscale{1.0}
\plotone{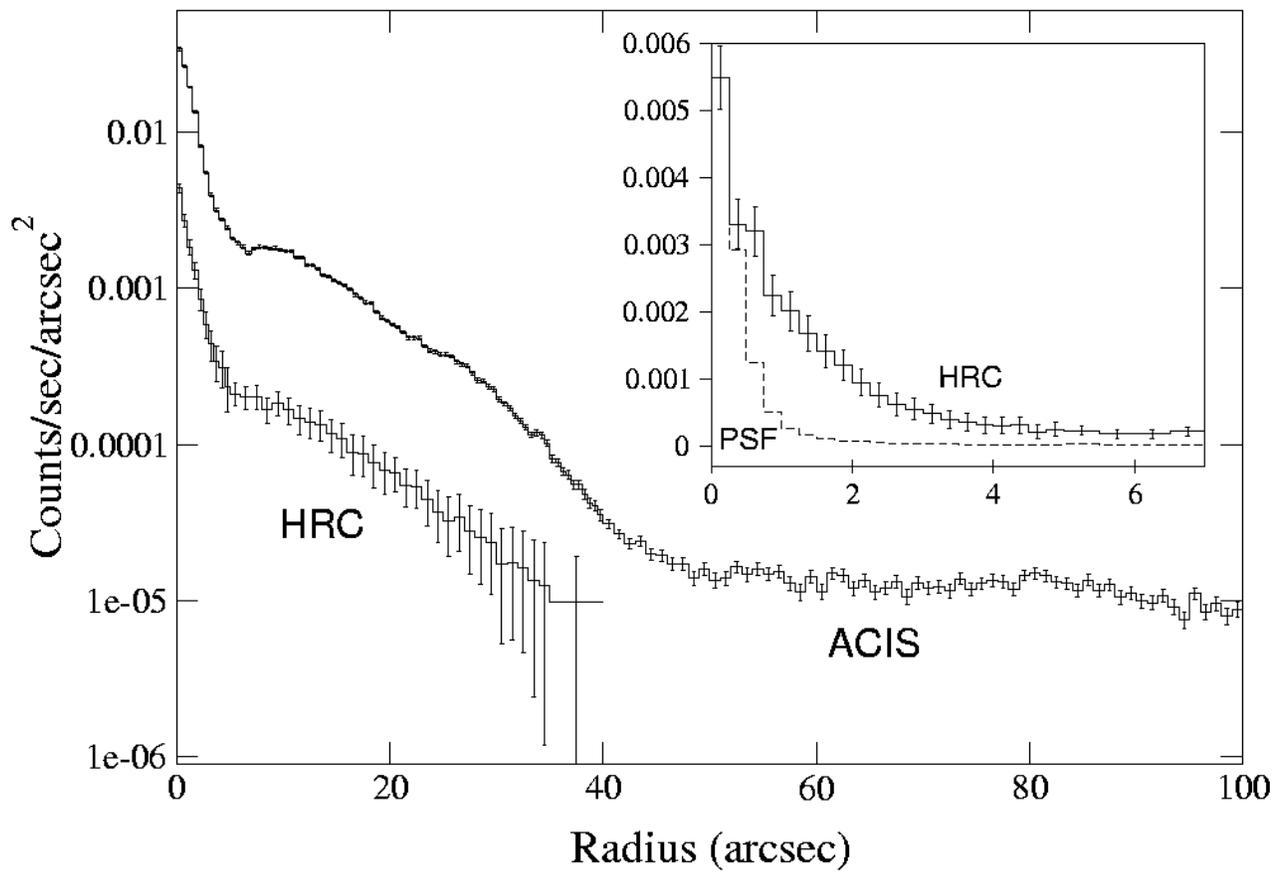}
\caption{The radial profile of \g21 \ using HRC and ACIS.
The HRC profile extends out only to a radius $\sim$ 40\as \ since
the background dominates at larger radii.
The inset shows the HRC profile with 0\as.25 bins and the
expected point-spread function (PSF, dashed line). }
\end{figure}

\clearpage

\begin{figure}[tb]
\epsscale{0.75}
\plotone{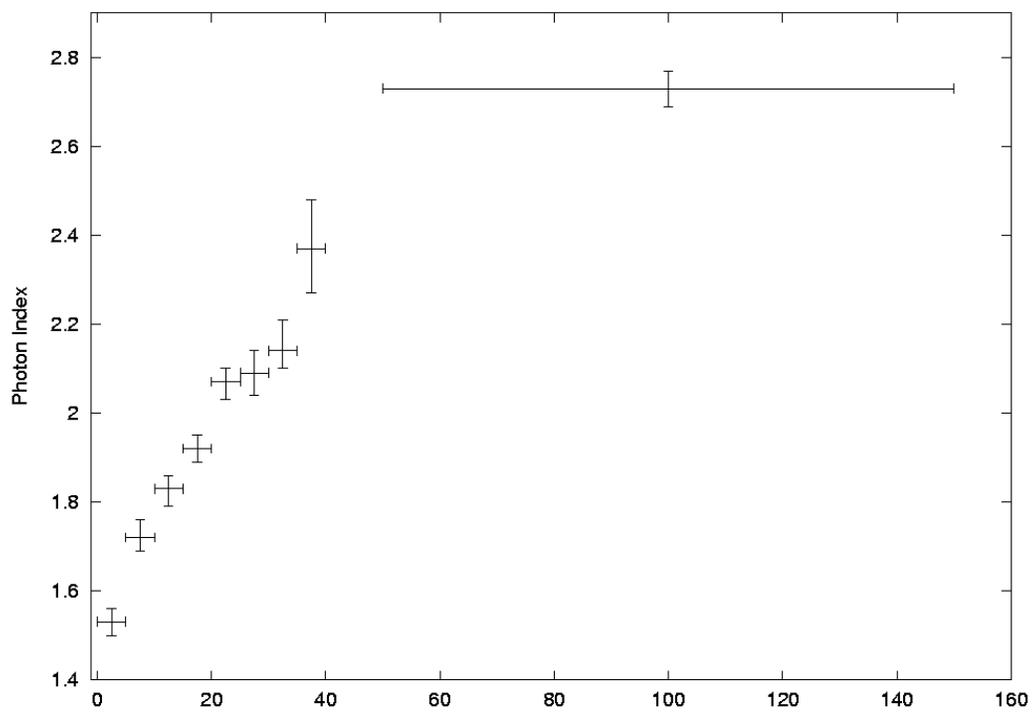}
\caption{The Power Law photon index $\Gamma$ as a function of radius 
(in arcseconds).
Zero radius corresponds to the brightest emission region of the plerion,
centered at $RA$ =
18$^{\rm h}$33$^{\rm m}$33$^{\rm s}\!.$50, $Dec$ =
$-$10$^\circ$34$^\prime$6$^{\prime\prime}\!.$6 (J2000). 
We show the power law index of the inner 40\as \ radius core
(in radial increments of 5\as) and the extended component
 (50\as--150\as), for $N_H$ = 2.24 $\times$ 10$^{22}$ cm$^{-2}$. }
\epsscale{1.0}
\end{figure}

\clearpage

\begin{figure}
\epsscale{1.0}
\plottwo{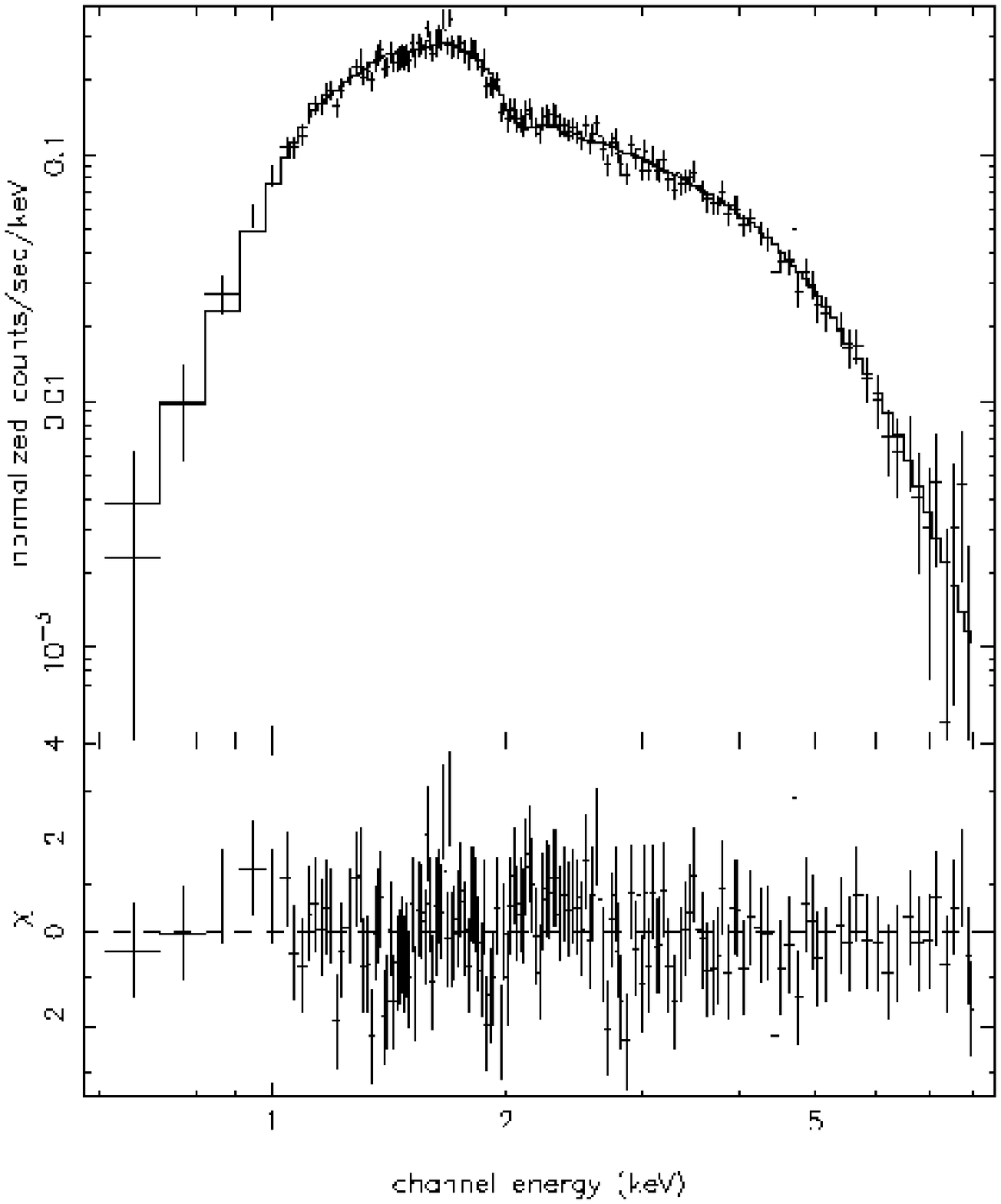}{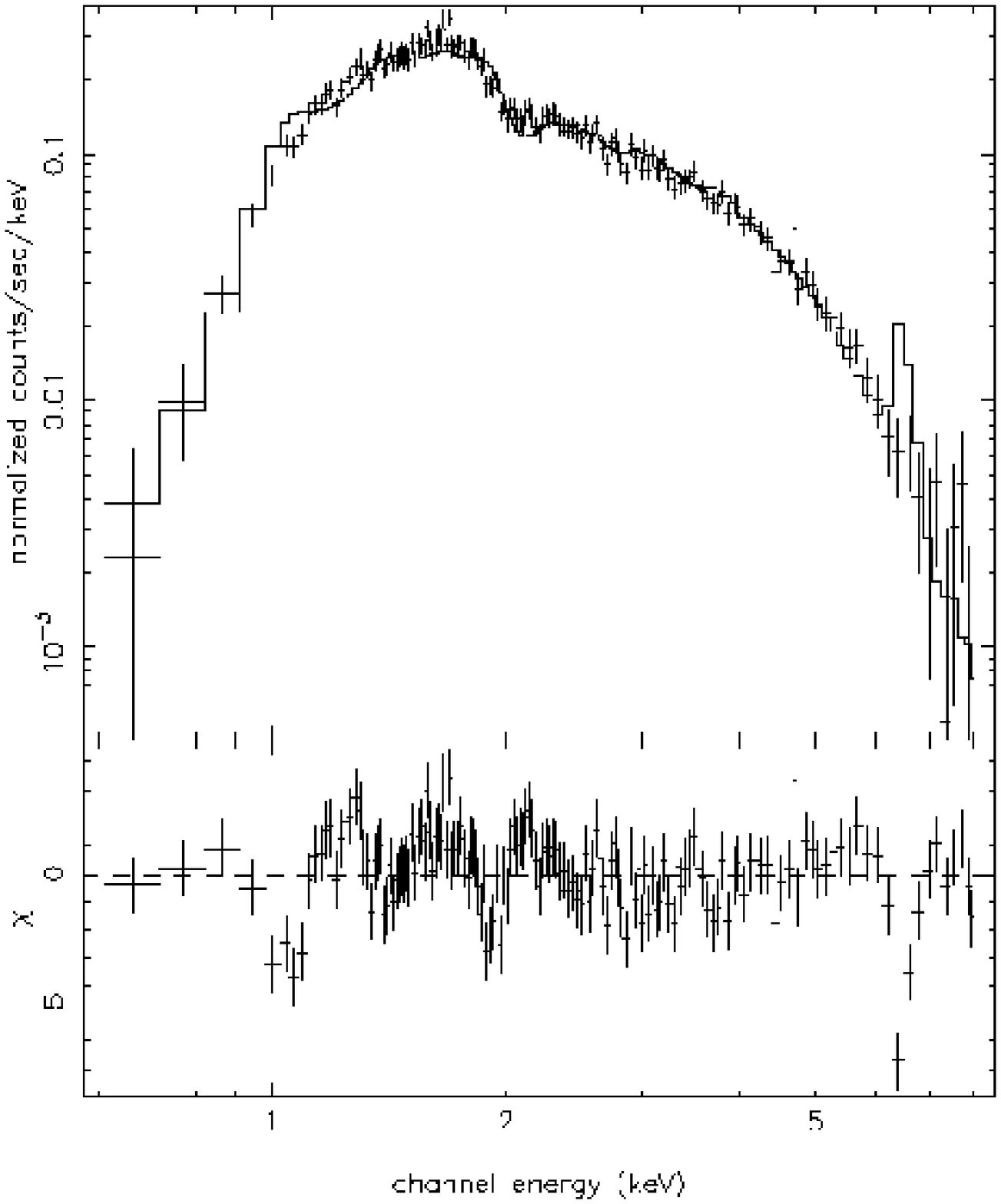}
\epsscale{1.0}
\caption{The spectra of the extended component (50\as--150\as)
fitted with a power law model (left) and the $\it{APEC}$ thermal model
assuming solar abundances (right).
We show that while a non-thermal model fits this component
adequately, the thermal model fails.
The parameters of the best fit (power law) are summarized in Table~1.}
\end{figure}
             
\clearpage

\begin{figure}[tb]
\epsscale{0.75}
\plotone{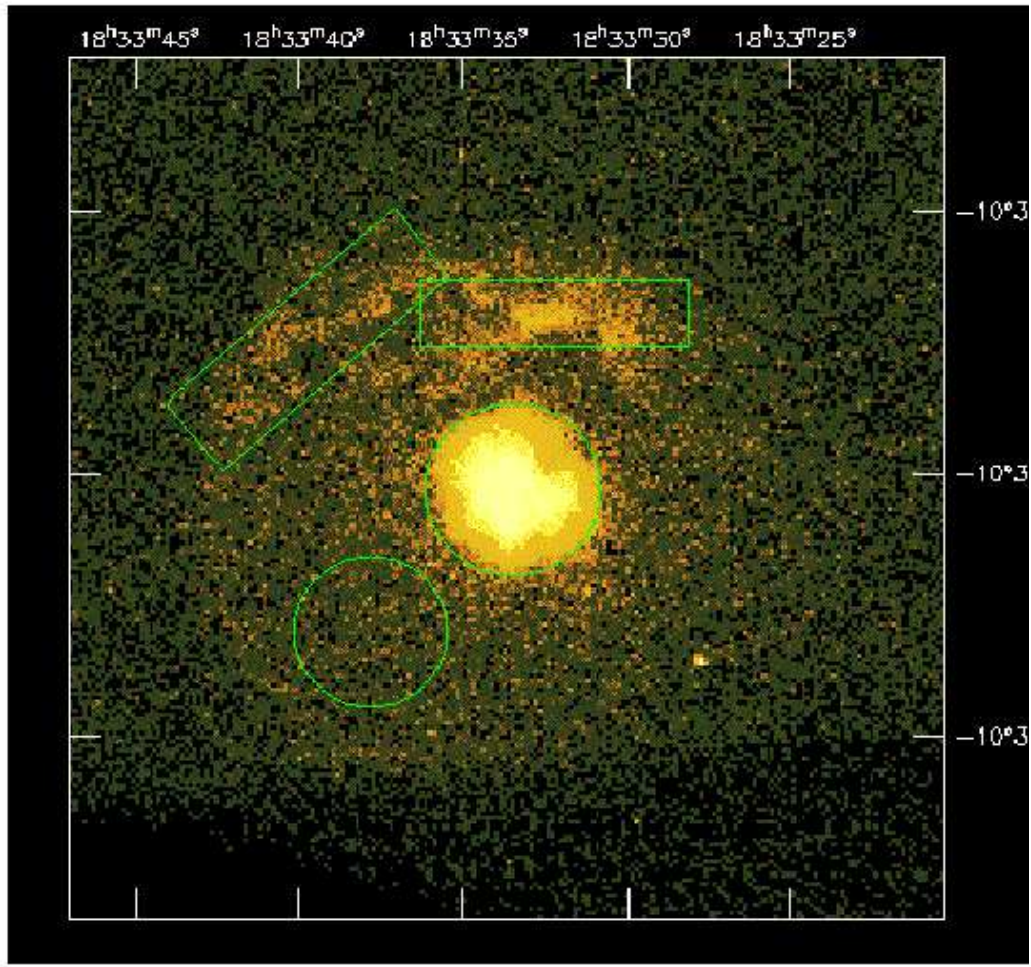}
\epsscale{1.00}
\caption{
The \chandra \ ACIS-S (0.5-10~keV) image of
G21.5-0.9 showing the inner core ($R$$\sim$40\as)
and the diffuse component extending out to
a radius of $R$$\sim$2\am.5.
We also show (in green) the regions selected for the spectral analysis:
the north-eastern knots, the brightest north-western edge,
and the south-eastern region.}
\end{figure}      

\clearpage

\begin{figure}
\epsscale{0.5}
\plotone{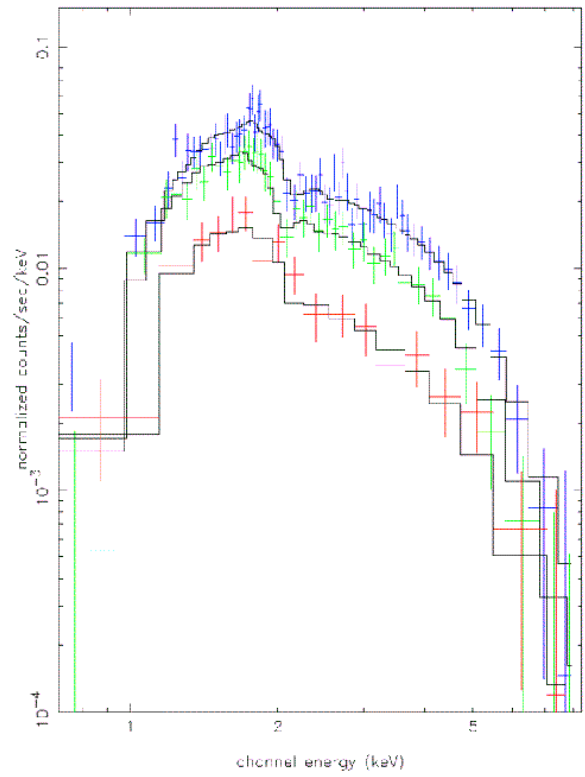}
\epsscale{1.0}
\caption{
The data and the
 fitted power law models for selected regions from the extended
component, shown in Fig.~8. The blue, green, and red colors
correspond to the bright north-eastern knots, the north-western edge, 
and the south-eastern region, respectively.}
\end{figure}

\clearpage

\begin{figure}[tb]
\epsscale{0.4}
\plotone{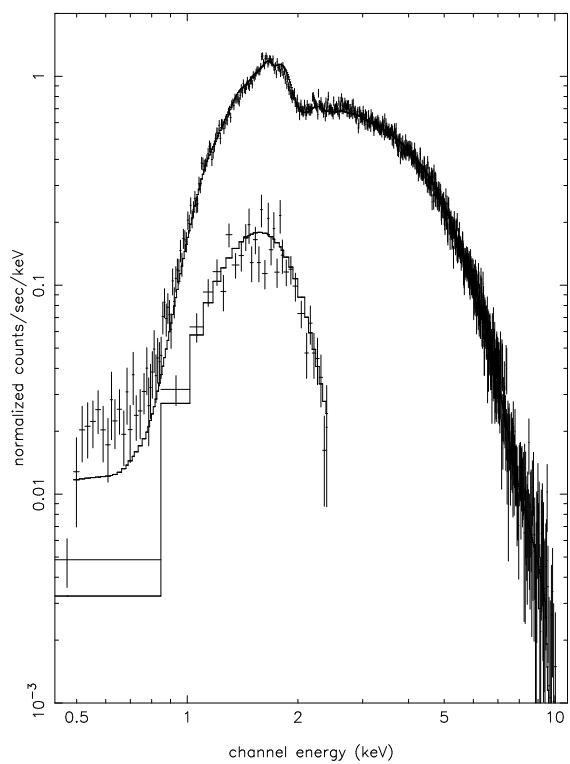}
\epsscale{1.0}
\caption{The PSPC and ACIS-S3 spectra of the inner 40\as \ core
 fitted with a power law model (\S 4.2).}
\epsscale{1.0}
\end{figure}

\clearpage

\begin{figure}[tb]
\epsscale{0.8}
\plottwo{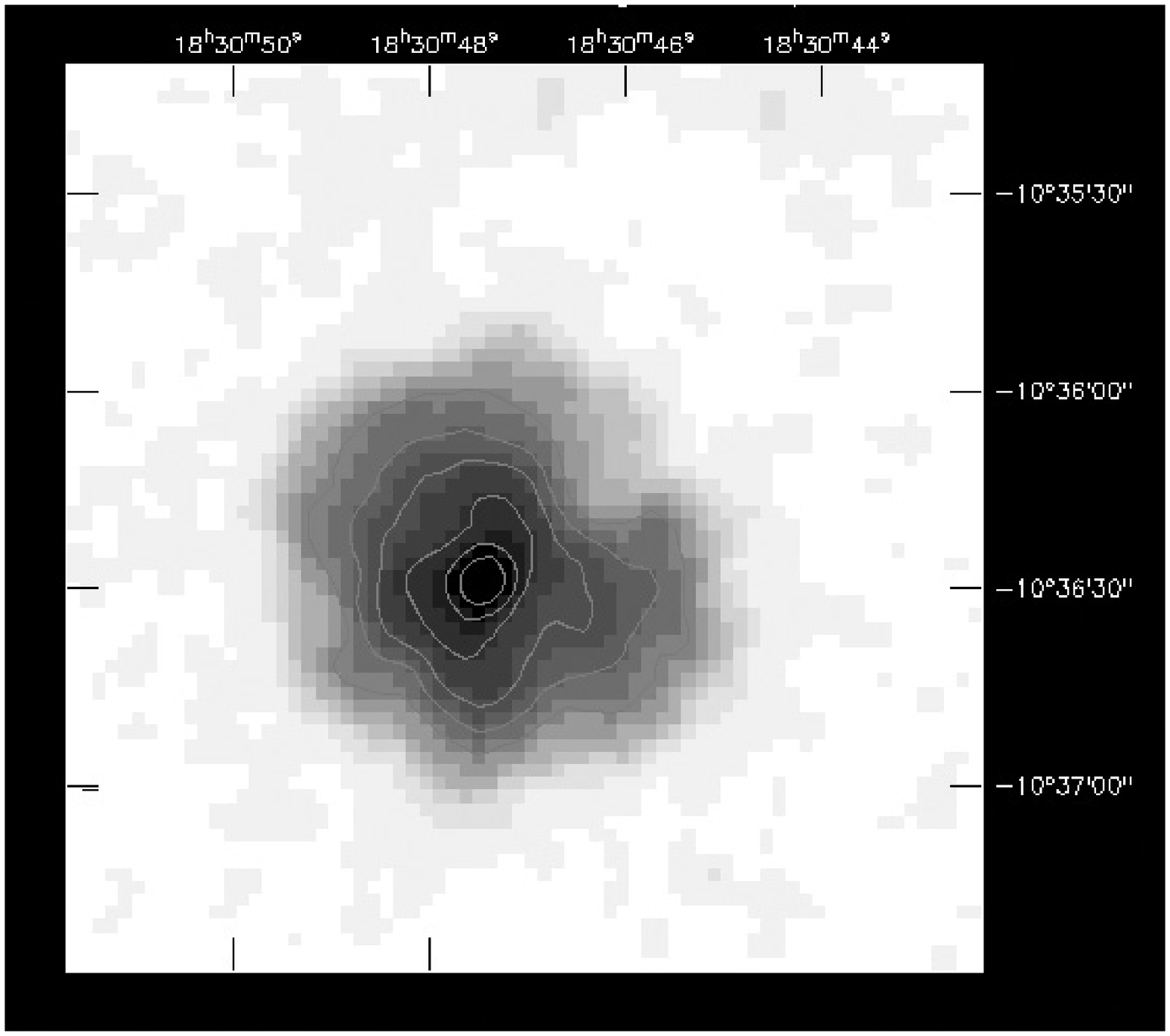}{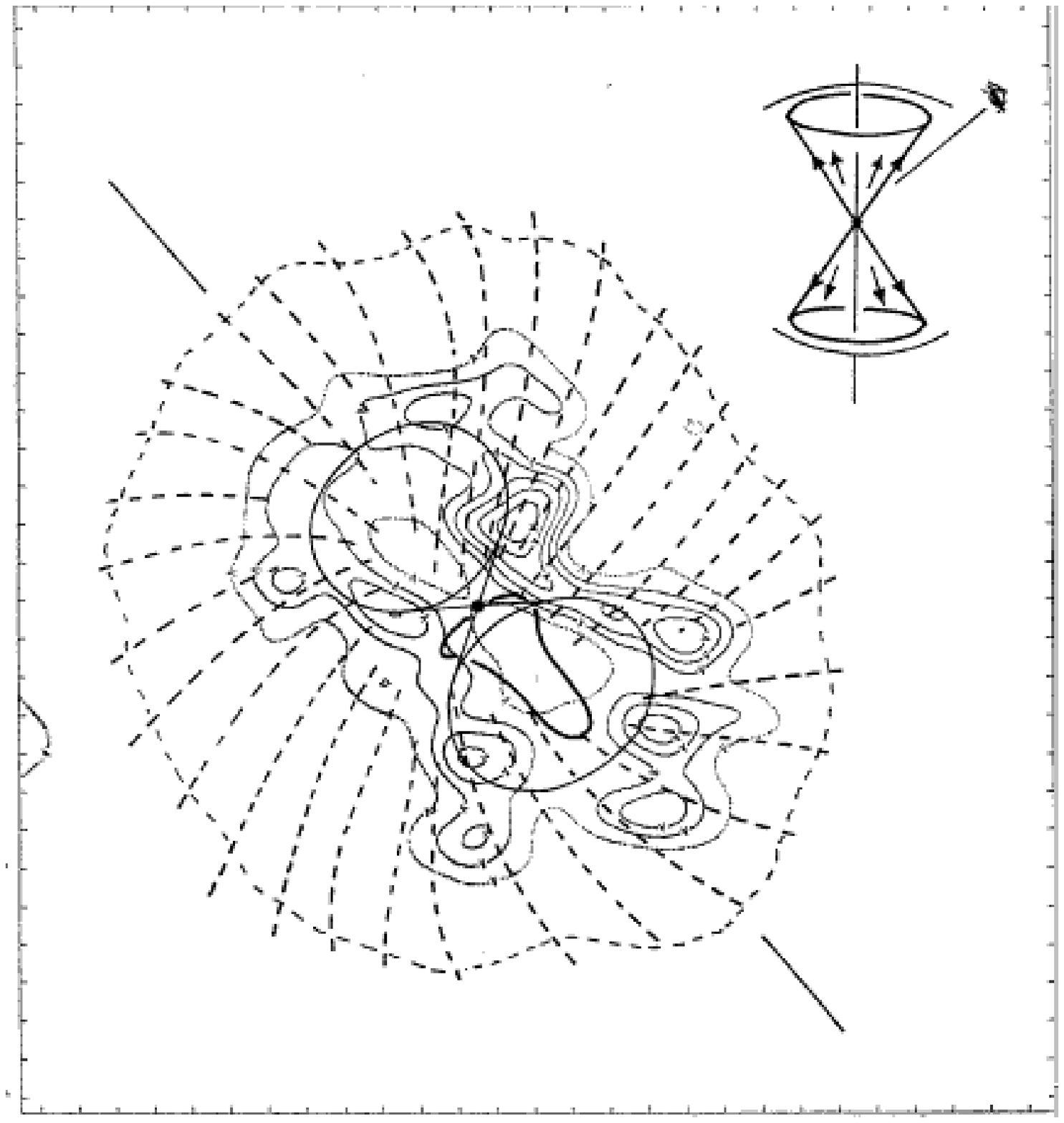}
\epsscale{1.0}
\caption{The ACIS-S and 22.3 GHz images of G21.5-0.9, binned to the
same scale (B1950 coordinates). The morphology of the plerion in X-rays follows
the enhancements seen in the filamentary radio structures.}
\end{figure}

\clearpage

\begin{table}
\begin{center}
\caption{Spectral parameters of the fits to the low-surface-brightness
extended component using ACIS-S3 observations.
The ranges are at the 90\% confidence level.}

\begin{tabular}{cccc|c}
\tableline\tableline

Model &  $N_H$\tablenotemark{a} ($\times$10$^{22}$ cm$^{-2}$)
 & Model Parameter & Norm. ($\times$10$^{-3}$) &
 $\chi^2_{\nu}$ ($\nu$) \\
\tableline

 Power Law & 1.83 (1.77--1.90) & $\Gamma$ =  2.36 (2.30-2.43) & 4.36 (4.04--4.78) & 1.0 (658) \\
 APEC\tablenotemark{b} & 1.63 (1.58--1.69) & $kT$ = 3.33 (3.14--3.54) & 8.22 (7.89--8.59) & 1.15 \\
 ${\it PSHOCK}$ & 1.53 (1.48--1.57) & $kT$ = 3.50 (3.28--3.74) keV & 9.72 (9.23--10.2) & 0.86 (657) \\
             &                   & $\tau$\tablenotemark{c} = 1.1 (0.8--1.3)$\times$10$^9$ cm$^{-3}$~s & & \\ \tableline
\end{tabular}

\tablenotetext{a}{Best fit values are indicated. When freezing $N_H$ to 2.2~$\times$
10$^{22}$~cm$^{-2}$, the power law model gives $\Gamma$ = 2.73~$\pm$~0.04 
($\chi^2_{\nu}$~=~0.98; $\nu$~=~659),
and thermal models are rejected ($\chi^2_{\nu}$$\sim$1.5)}
\tablenotetext{b}{Assuming solar abundances (http://hea-www.harvard.edu/APEC)} 
\tablenotetext{c}{The ionization timescale, $n_et$; where $n_e$ is
 the postshock electron density, and
$t$ is the age of the shock}
\end{center}
\end{table}

\clearpage             

\begin{table}
\begin{center}
\caption{{\it Chandra}/HRC observational sets}
\begin{tabular}{cccccccc}
\tableline\tableline
ObsID\tablenotemark{a}      & Detector      & Date          & Epoch\tablenotemark{b}     &
 $T_{\it{span}}$\tablenotemark{c}     & $T_{\it{exp}}$\tablenotemark{d}  & $SIM_X$\tablenotemark{e}    & $N$\tablenotemark{f} \\
         &              &               & MJD           & s                     & s
   & mm         & \\
\tableline
{\it 143\/}      & HRC-I        & Sep 4, 1999   & 51425.339     & 15 243                & 15 243
        & $-$1.0160     & 179 \\
{\it 1242\/}     & HRC-I        & Sep 4, 1999   & 51425.524     & 14 855                & 14 855
        & $-$1.0160     & 555 \\
{\it 147\/}      & HRC-S        & Sep 5, 1999   & 51426.625     & 9 747                 & 9 747
        & $-$1.3705     & 496 \\
{\it 1406\/}$^a$        & HRC-I         & Oct 25, 1999  &               & 27 572                & 25
621             & $-$1.0389     & \\
{\it 146\/}      & HRC-S        & Feb 16, 2000  & 51591.026     & 10 334                & 10 265
        & $-$1.4282     & 518 \\
\tableline
{\it 1406$^a$\/}   &              &               & 51476.316     & 13 797                &
        &               & 543 \\
{\it 1406$^b$\/}    &              &               & 51476.475     & 13 726                &
        &               & 590 \\
\tableline
\end{tabular}

\tablenotetext{a}{the ObsID {\it 1406\/} set was divided in two parts, see explanations in Section 5}
\tablenotetext{b}{epoch refers to the central time of the exposure}
\tablenotetext{c}{time span}
\tablenotetext{d}{effective exposure time}
\tablenotetext{e}{focus position along the telescope axis;
 the optimal positions are $-1.04$ mm for HRC-I, and $-1.53$ mm for HRC-S}
\tablenotetext{f}{number of counts used in the analysis}
\label{t:hrc_sets}
\end{center}
\end{table}                        

\clearpage

\begin{table}
\begin{center}
\caption{Comparison of the parameters of
the putative pulsar in \g21 \ with the Crab pulsar and the putative
pulsar in 3C~58 (Torii \ea \ 2000).}
\begin{tabular}{c|ccc|}
\tableline\tableline
        &       \g21    &       Crab & 3C~58 \\ \tableline
Distance (kpc) &       5       &       2  & 3.2 \\
$L_X$\tablenotemark{a} (10$^{35}$ erg~s$^{-1}$) &        3.3$D_5^2$      & 210 & 0.2  \\
$\dot{E}$ (10$^{37}$ erg~s$^{-1}$) &
3--6    &  47 & 0.4 \\
$\sigma$\tablenotemark{b} &      (4--11)$\times$10$^{-4}$   &     3$\times$10$^{-3}$
& $\sim$ 6$\times$10$^{-3}$ \\
$P$ (ms)\tablenotemark{c} &
$144 \tau_3^{-0.5} \dot{E}_{37}^{-0.5}$ & 33 & 440\\
$B_0$ ($10^{13}$ Gauss) & $1.1~\tau_3^{-1}\dot{E}_{37}^{-0.5}$      & 0.4 & 6 \\
\tableline
\end{tabular}

\tablenotetext{a}{X-ray luminosity of the plerion in the 0.5--10 keV range}
\tablenotetext{b}{Pulsar wind parameter defined as the ratio of
the Poynting flux to the
kinetic energy flux (Kennel and Coroniti 1984)}
\tablenotetext{c}{
The following notations are used: $\tau_3$ is the age of the SNR in units of
3 kyr; $\dot{E}_{37}$ is the spin-down energy
loss in units of 10$^{37}$ erg~s$^{-1}$, $D_5$ is the distance in units
of 5 kpc}
\end{center}
\end{table}

\clearpage       


\clearpage


\begin{thebibliography}{}

\bibitem[Allen et al.(1999)] {All99} Allen, G. E., Gotthelf, E. V., \& Petre, R.
1999, astro-ph/9908209

\bibitem[Asaoka(1990)]{asa90} Asaoka, I., \& Koyama, K. 1990, PASJ, 42, 625

\bibitem[Becker(1992)]{bec92} Becker, P. A. 1992, \apj, 397, 88

\bibitem[Becker and Tr\"umper(1997)]{bec97} Becker, W. \& Tr\"umper, J. 1997,
A\&A, 326, 682 

\bibitem[Becker and Szymkowiak(1981)]{bec81}
Becker, R. H. \& Szymkowiak, A. E. 1981,
248, L23

\bibitem[Bocchino et al.(2001)]{boc01} Bocchino, F., Warwick, R. S., 
Marty, P., Lumb, D., Becker, W., \&  Pigot, C. 2001, A\&A, 369, 1078

\bibitem[Borkowski et al.(2000)]{bor00} Borkowski, K. J., Lyerly, W. J., 
\& Reynolds, S. P. 2000, \apj, 548, 820 

\bibitem[Buccheri et al.(1983)]{buc83}
Buccheri, R., et al. 1983, A\&A, 128, 245

\bibitem[Chevalier(2000)]{che00} Chevalier, R. A. 2000, ApJ, 539, L45

\bibitem[Dyer et al.(2000)]{kri00} Dyer, K. K., Reynolds, S. P., Borkowski, K. J.,
\& Petre, R. 2000, \apj, 551, 439 

\bibitem[Davelaar(1986)]{dav86} Davelaar, J., \& Smith, A. B, \& Becker,
R. H. 1986, ApJ, 300, L59

\bibitem[F\"urst et al.(1988)]{fur88}
F\"urst, E., Handa, T., Morita, K., Reich, P., Reich, W., Sofue, Y. 1988,
PASJ, 40, 347

\bibitem[Gallant and Tuffs(1998)]{gal98} Gallant, Y. A., \& Tuffs, R. J. 1998,
Mem. Soc. Astron. Ital., 69, 963 

\bibitem[Green(2000)] {gre00} Green, D. A. 2000, 
http://www.mrao.cam.ac.uk/
 
\bibitem[Kaspi et al.(1996)]{kas96} Kaspi, V. M., Manchester, R. N., Johnston, S.,
Lyne, A. G., D'amico, N. 1996, AJ,
 111 (5), 2028

\bibitem[Kennel and Coroniti(1984)]{ken84} Kennel, C. F. \& Coroniti, F. V.
1984, ApJ, 283, 710 (KC~84)

\bibitem[Kotsevich and Pavlov(2001)]{kop01} 
Koptsevich, A.~B., \& Pavlov, G.~G. 2001, ApJ, in preparation

\bibitem[Lucke(1978)]{luc78} Lucke, P. B. 1978, A\&A, 64, 367
 
\bibitem[Morsi(1987)]{mor87} Morsi, H. W. and Reich, W. 1987, A\&AS,
69, 533

\bibitem[\"Ogelman(1995)]{oge95} \"Ogelman, H. B. 1995, in `The Lives of
Neutron Stars', Eds. M.~A.~Alpar, \"U.~Kiziloglu, \& J. van Paradijs, 
NATO ASI series, P.~116

\bibitem[Pavlov, Zavlin, and Tr\"umper(1999)]{pav99}
Pavlov, G. G., Zavlin, V. E., \& Tr\"umper, J. 1999, ApJ, 511, L45

\bibitem[Predehl and Schmitt(1995)]{pre95} Predehl, P. and
Schmitt, J.~H.~M.~M. 1995,  A\&A, 293, 889

\bibitem[Reynolds and Aller(1985)]{rey85} Reynolds, S. P., \& Aller 1985,
AJ, 90, 2312

\bibitem[Reynolds and Chevalier(1981)]{rey85} Reynolds, S. P. \&
Chevalier, R. A. 1981, ApJ, 245, 912

\bibitem[Reynolds and Keohane(1999)]{rey99} Reynolds, S. P., \& Keohane, J. W.
1999, ApJ, 525

\bibitem[Salter et al.(1989)]{sal89} Salter, C. \ea \ 1989, ApJ, 338,
171

\bibitem[Salvati et al. (1998)]{sal98} Salvati, M., Bandiera, R., Pacini, F, \& Woltjer, L.
1998, Mem. Soc. Astron. Ital., 69, 1023

\bibitem[Seward and Wang(1988)]{sew88} 
  Seward, F. D. and Wang, Z. 1988, ApJ, 332, 199    

\bibitem[Slane et al.(2000)]{sla00} Slane P.~O. \ea \ 2000, ApJ, 533, L29

\bibitem[Torii, K. et al.(2000)]{tor00} Torii, K., Slane, P. O., Kinugasa, K., Hashimotodani, K., and Tsunemi, H. 2000, PASJ, 52, 875, astro-ph/0006034

\bibitem[Warwick et al.(2000)]{war00} Warwick, R. S. \ea \  2000, astro-ph/0011245

\bibitem[Weisskopf et al.(1996)]{wei96} Weisskopf, M. C., O'dell, S. L., and van Speybroeck, L. P.
1996, Spie, 2805, 2

\bibitem[Willingale et al.(1991)]{wil91} Willingale, R., Aschenbach, B., Griffiths, R. G.,
 Sembay, S., Warwick, R. S., Becker, W.,
 Abbey, A. F., Bonnet-Bidaud, J.-M. 1991, A\&A, 365, L212


\bibitem[Woltjer et al.(1997)]{wol97} Woltjer, L., Salvati, M., Paciti, F., and Bandiera, R., 1997,
A\& A, 325, 295

\end{thebibliography}
\end{document}